\documentclass[acmsmall]{acmart}

\setcopyright{cc}
\setcctype{by}
\acmDOI{10.1145/3808324}
\acmYear{2026}
\acmJournal{PACMPL}
\acmVolume{10}
\acmNumber{PLDI}
\acmArticle{246}
\acmMonth{6}
\acmSubmissionID{pldi26main-p537-p}
\received{2025-11-14}
\received[accepted]{2026-04-03}

\usepackage[T1]{fontenc}
\usepackage{enumitem}
\usepackage{xspace}
\usepackage[linesnumbered,vlined]{algorithm2e}
\usepackage[table]{xcolor}
\usepackage{wrapfig}
\usepackage{listings}
\usepackage{multirow}
\usepackage{pifont}
\usepackage{setspace}
\usepackage{hyperref}
\usepackage{pgfplots}
\pgfplotsset{compat=1.18}
\usepackage{fancyvrb}

\usepackage{tikz}
\DeclareRobustCommand{\circled}[1]{%
  \tikz[baseline=(char.base)]{
    \node[shape=circle,draw,inner sep=0.5pt,minimum size=7pt] (char)
    {\bfseries\scriptsize #1};
  }%
}

\makeatletter
\let\old@lstKV@SwitchCases\lstKV@SwitchCases
\def\lstKV@SwitchCases#1#2#3{}
\makeatother
\usepackage{lstlinebgrd}
\makeatletter
\let\lstKV@SwitchCases\old@lstKV@SwitchCases

\lst@Key{numbers}{none}{%
    \def\lst@PlaceNumber{\lst@linebgrd}%
    \lstKV@SwitchCases{#1}%
    {none:\\%
     left:\def\lst@PlaceNumber{\llap{\normalfont
                \lst@numberstyle{\thelstnumber}\kern\lst@numbersep}\lst@linebgrd}\\%
     right:\def\lst@PlaceNumber{\rlap{\normalfont
                \kern\linewidth \kern\lst@numbersep
                \lst@numberstyle{\thelstnumber}}\lst@linebgrd}%
    }{\PackageError{Listings}{Numbers #1 unknown}\@ehc}}
\makeatother

\AtBeginDocument{%
  }

\newcommand{\cmark}[0]{\color{green!50!black}{\ding{51}}}
\newcommand{\xmark}[0]{\color{red!50!black}{\ding{55}}}
\newcommand{\nmark}[0]{\color{yellow!40!black}{\raisebox{0.7ex}{\bf\Large\texttildelow}}}
\newcommand{\ignore}[1]{}

\newcommand{\tool}{\textsc{LibFpp}\xspace}

\newcommand{\myparagraph}[1]{\vspace{0.35em}\noindent\emph{#1.}}
\newcommand{\rn}[1]{\expandafter{\romannumeral #1\relax}}

\theoremstyle{definition}

\renewcommand{\equationautorefname}{\relax}
\def\equationautorefname#1{}

\definecolor{mycolor}{rgb}{0.122, 0.435, 0.698}
\definecolor{darkgreen}{rgb}{0.0, 0.435, 0.0}

\usepackage{tcolorbox}
\newcommand{\result}[1]{%
\begin{tcolorbox}[colframe=blue,boxrule=0.5pt,arc=4pt,left=6pt,right=6pt,top=6pt,bottom=0pt,boxsep=0pt,width=\columnwidth]%
      {#1}
\end{tcolorbox}%
}

\begin{document}

\title{Persistent Iterators with Value Semantics}

\author{Yihe Li}
\orcid{0009-0008-2257-0406}
\affiliation{%
  \institution{National University of Singapore}
  \city{Singapore}
  \country{Singapore}
}
\email{yihe.li@u.nus.edu}

\author{Gregory J. Duck}
\orcid{0000-0002-0837-9671}
\affiliation{%
  \institution{National University of Singapore}
  \city{Singapore}
  \country{Singapore}
}
\email{gregory@comp.nus.edu.sg}



\begin{abstract}
{\em Iterators} are a fundamental programming abstraction for traversing and modifying elements in containers in mainstream imperative languages such as \texttt{C++}.
Iterators provide a uniform access mechanism that hides low-level implementation details of the underlying data structure.
However, iterators over {\em mutable} containers suffer from well-known hazards including invalidation, aliasing, data races, and subtle side effects.
{\em Immutable} data structures, as used in functional programming languages, avoid the pitfalls of mutation but rely on a very different programming model based on recursion and higher-order combinators (\texttt{map}, \texttt{foldl}, \texttt{traverse}, etc.) rather than iteration.
However, these combinators are not always well-suited to expressing certain algorithms, and recursion can expose implementation details of the underlying data structure.

In this paper, we propose {\em persistent iterators}---a new abstraction that reconciles the familiar iterator-based programming style of imperative languages with the semantics of persistent data structures.
A persistent iterator snapshots the version of its underlying container at creation, ensuring safety against invalidation and aliasing.
Iterator operations (\texttt{++}, \texttt{erase}, etc.) operate on the iterator-local copy of the container, giving true value semantics: variables can be rebound to new persistent values while previous versions remain accessible.
We implement our approach in the form of \tool---a \texttt{C++} container library providing persistent vectors, maps, sets, strings, and other abstractions as persistent counterparts to the {\em Standard Template Library} (STL).
Our evaluation shows that \tool retains the expressiveness of iterator-based programming, eliminates iterator-invalidation, and achieves asymptotic complexities comparable to STL implementations---albeit with the higher constant-factor overheads of persistence.
Our design targets use cases where persistence and safety are desired, while allowing developers to retain familiar iterator-based programming patterns.
\end{abstract}

\maketitle

\section{Introduction}

{\em Iterators} are one of the most ubiquitous and useful abstractions among programming languages and form the backbone of the standard libraries for popular languages like \verb_C++_~\cite{cpp23}, Python~\cite{python_iterators}, and Rust~\cite{rust_iterators}.
Acting as {\em generalized pointers} into structured data, iterators provide a uniform and representation-independent interface for accessing and manipulating container elements.
This uniformity makes iterators general in principle: traversal logic can be written once and reused across a wide range of data structures and algorithmic contexts.
Such uniformity enables iterators to become the bases of generic algorithms such as sorting and searching, eventually leading to the birth of powerful and generic libraries like the {\em Standard Template Library} (STL) of \verb_C++_~\cite{stepanov1995stl, cpp23} and Java's {\em Collections Framework}~\cite{java_collections_framework}.
However, when coupled with {\em mutation}, iterators can be the source of a well-known class of subtle bugs in programs~\cite{cpp23, lakos_large_scale_cpp}.
When the underlying structure is modified, iterators may become {\em invalidated}, possibly leading to vulnerabilities (\verb_C++_), runtime exceptions (Java/C\#), or undefined/unpredictable behavior (Python).
For example, the \verb_C++_ STL treats iterator invalidation as an example of {\em undefined behavior} (UB), often leading to {\em use-after-free} vulnerabilities, and each modifying operation tends to have its own specific invalidation rules.
Furthermore, if the container is modified concurrently through an active iterator, consistency issues may arise without careful synchronization of observers~\cite{Nikola15}.

Other programming languages do not typically expose an iterator abstraction.
For example, {\em functional programming} (FP) languages, such as Haskell, use {\em immutable}/{\em persistent} data structures that cannot be mutated after construction~\cite{Driscoll86}.
Under this setting, modifying operations can only create {\em new} versions of existing data structures without destroying the old copy.
For example, Haskell's \verb+Data.Map.insert+ creates a new associative map without destroying the original, meaning that {\em both} the new and original map values can be used after the operation.
Note that updates to persistent data structures typically do {\em not} require a deep copy of the entire data structure.
Rather, only the specific nodes that require updating are copied, and the rest of the structure is {\em shared} with the original---a.k.a. {\em structural sharing}, which is essential for an efficient (both time and memory) implementation of persistence.
The persistent programming model preempts the invalidation-via-mutation problem: each observer operates on their own local copy of the data structure without interference from concurrent updates.
Prior work demonstrates that persistent data structures can achieve favorable asymptotic efficiency and practical performance for many workloads~\cite{Okasaki95}, albeit with nontrivial constant-factor overheads compared to their imperative counterparts.

Yet, existing persistent data structure implementations tend not to expose a general iterator abstraction.
One key problem is the fundamental difference between {\em immutable} data structures and {\em mutable} references used by iterators, which are often viewed as incompatible programming paradigms.
Instead, Haskell's \verb+Data.Map+ exposes a collection of higher-order {\em combinators} (\verb+map+, \verb+filter+, \verb+foldl+, \verb+traverse+, etc.) for data-structure traversal and modification.
Such combinators capture many common traversal/transformation patterns, and can be {\em composed}, allowing for the construction of complex algorithms.
However, while combinators capture many traversal patterns, they are fixed in arity and structure, and thus may be less natural or less idiomatic for expressing algorithms requiring stateful or early-exit iteration.
For Haskell lists, one valid option is to fall back to using {\em recursion} instead of combinators.
However, this option is rarely exposed by existing {\em opaque} data structure APIs, such as \verb+Data.Map+, primarily for performance reasons.
In contrast, \verb_C++_-style iterators are both highly expressive and abstract.

In this paper, we bridge the gap between the two seemingly incompatible programming pa\-ra\-digms in the form of {\em persistent iterators}---an iterator-centric abstraction combined with the safety of persistent data structures.
Unlike read-only {\em const iterators}, which deny updates via iterators altogether, persistent iterators still support updates by working over a {\em local copy} of the underlying data structure rather than the original via a mutable reference.
This means that any update via the iterator (e.g., assignment, insertion, erasure) does not invalidate (or even modify) the original container, nor any other local copy from any other iterator.
Essentially, persistent iterators support STL-style iterator operations---including updates via non-const iterators---while maintaining immutability guarantees.
We show that persistent iterators retain much of the expressive power of traditional mutable iterators, as well as providing a uniform abstraction that does not expose implementation details of the underlying structure.
Furthermore, persistent iterators are immune to iterator invalidation---thereby eliminating a common class of programming bug.

\lstdefinestyle{CStyle}{
  language=C,
  basicstyle=\ttfamily\scriptsize,
  keywordstyle=\color{blue},
  morekeywords={string, iterator, list, char32_t},
  keywordstyle=[2]\color{teal},
  commentstyle=\color{gray}\itshape,
  stringstyle=\color{red},
  showstringspaces=false,
  numbers=none,
  escapeinside={(*@}{@*)},
  tabsize=2,
  xleftmargin=1em,
  frame=single,
  framesep=2pt,
  backgroundcolor=\color{gray!8},
}

\lstdefinestyle{HStyle}{
  language=haskell,
  basicstyle=\ttfamily\scriptsize,
  keywordstyle=\color{blue},
  morekeywords={Seq},
  keywordstyle=[2]\color{teal},
  commentstyle=\color{gray}\itshape,
  stringstyle=\color{red},
  showstringspaces=false,
  numbers=none,
  escapeinside={(*@}{@*)},
  tabsize=2,
  xleftmargin=1em,
  frame=single,
  framesep=2pt,
  backgroundcolor=\color{gray!8},
}

\begin{figure}
\centering
{\footnotesize

\begin{tabular}{c}
\begin{minipage}{0.47\linewidth}
\centering
\lstset{style=CStyle, backgroundcolor=\color{green!8}}
\begin{lstlisting}
string filterAscii(string s) {
  string::iterator i;
  for (i = s.begin(); i != s.end(); )
    if (*i >= 0x7F) i.erase();
    else            ++i;
  return i.value();
}
\end{lstlisting}
\end{minipage}
\\
\textcolor{red!60!black}{(a)} \texttt{C++} \tool \texttt{string}
\\\\
\begin{minipage}{0.47\linewidth}
\centering
\lstset{style=CStyle, backgroundcolor=\color{red!8}}
\begin{lstlisting}
void filterAscii(std::list<char32_t> &s) {
  for (auto i = s.begin(); i != s.end(); )
    if (*i >= 0x7F) i = s.erase(i);
    else            ++i;
}
\end{lstlisting}
\end{minipage}
\\
\textcolor{red!60!black}{(b)} \texttt{C++} STL \texttt{std::list<char>}
\\\\
\begin{minipage}{0.47\linewidth}
\centering
\lstset{style=CStyle, backgroundcolor=\color{red!8}}
\begin{lstlisting}
void filterAscii(std::string &s) {
  std::string t;
  for (auto c: s) if (c < 0x7F) t += c;
  s.swap(t);
}
\end{lstlisting}
\end{minipage}
\\
\textcolor{red!60!black}{(c)} \texttt{C++} STL \texttt{std::string}
\end{tabular}
\begin{tabular}{c}
\begin{minipage}{0.46\linewidth}
\centering
\lstset{style=HStyle, backgroundcolor=\color{blue!8}}
\begin{lstlisting}
filterAscii :: String -> String
filterAscii = filter (\c -> ord c < 0x7F)
\end{lstlisting}
\end{minipage}
\\
\textcolor{red!60!black}{(d)} Haskell \texttt{String} (combinator)
\\\\
\begin{minipage}{0.46\linewidth}
\centering
\lstset{style=HStyle, backgroundcolor=\color{blue!8}}
\begin{lstlisting}
filterAscii :: String -> String
filterAscii [] = []
filterAscii (c:cs)
  | ord c < 0x7F = c : filterAscii cs
  | otherwise    = filterAscii cs
\end{lstlisting}
\end{minipage}
\\
\textcolor{red!60!black}{(e)} Haskell \texttt{String} (recursion)
\\\\
\begin{minipage}{0.46\linewidth}
\centering
\lstset{style=HStyle, backgroundcolor=\color{blue!8}}
\begin{lstlisting}
filterAscii :: Seq Char -> Seq Char
filterAscii s = go 0 s
  where
    go i s
      | i >= Seq.length s = s
      | otherwise =
          let c = Seq.index s i
          in if ord c >= 0x7F
             then go i (Seq.deleteAt i s)
             else go (i + 1) s
\end{lstlisting}
\end{minipage}
\\
\textcolor{red!60!black}{(f)} Haskell (\texttt{Seq Char})
\end{tabular}
}
\caption{Comparing a simple ``{\em filter non-ASCII characters from a string}'' function across paradigms using \texttt{C++} STL, \texttt{C++} \tool, and Haskell.
Version \textcolor{red!60!black}{(a)} is the idiomatic implementation using \tool persistent iterators.
Versions \textcolor{red!60!black}{(b)} and \textcolor{red!60!black}{(c)} are idiomatic implementations for \texttt{std::list<char>} and \texttt{std::string}, respectively.
Finally, we consider \textcolor{red!60!black}{(d)}/\textcolor{red!60!black}{(e)} as the idiomatic combinator/recursive implementation over Haskell \texttt{String}s, and \textcolor{red!60!black}{(f)} over Haskell (\texttt{Seq Char}) from \texttt{Data.Sequence} using explicit indices.\label{fig:examples}
}
\end{figure}

We have implemented and evaluated our approach in the form of \tool---a \verb_C++_ library that provides concrete implementations of common STL-style containers, such as {\em vectors}, {\em strings}, {\em sets}, and {\em maps}, in a persistent manner.
\tool is designed to resemble the \verb_C++_ STL in terms of functionality, algorithmic complexity, and API---where the main differences relate only to persistence and value semantics---reducing the learning curve for developers already familiar with STL-style programming.
\tool containers are built on top of {\em finger trees}~\cite{HINZ2006}---a persistent data structure that can be used to implement common STL container operations with comparable algorithmic complexity.
Persistent iterators are built on top of optimized {\em zippers}~\cite{HUET1997}---a persistent data structure for representing a ``pointer'' or ``cursor'' into another persistent data structure.
We show that zippers can be used to implement a rich STL-like iterator API, again with comparable algorithmic complexity.
We also show that, with sufficient optimization, \tool achieves the algorithmic and safety benefits of persistent containers and value semantics with practical efficiency consistent with the inherent constant factor overheads of persistence.

\subsection{Motivating Example}\label{sec:example}

As a motivating example, we consider a simple \verb+filterAscii+ function, which filters out all non-ASCII characters from a string.
Several different implementations over different string representations are shown in \autoref{fig:examples}.
A summary of the differences between different implementations is shown in \autoref{tab:filterAscii}.
Our core argument is that, despite \verb+filterAscii+ being a relatively trivial function, the STL \verb_C++_ and Haskell implementations do not simultaneously achieve all of the following properties:

\begin{itemize}[leftmargin=*]
\item[-] optimal $O(N)$ algorithmic {\em complexity};
\item[-] structural {\em sharing} (i.e., will the result share unmodified structure with the input?);
\item[-] {\em persistence} (i.e., is the data structure {\em not} mutated in-place?); and
\item[-] {\em abstraction} (i.e., are implementation details of the container exposed?).
\end{itemize}
Optimal {\em complexity} is desirable for performance,
structural {\em sharing} to minimize memory usage,
{\em persistence} for algorithmic and safety benefits, and
{\em abstraction} for modularity and elegance.

None of \autoref{fig:examples}~(b)-(f) achieves all four simultaneously.
For example, the optimal \verb_C++_ algorithm depends on the underlying string structure, with in-place erasure (\autoref{fig:examples}~(b)) for node-based containers \verb+std::list<char>+ and {\em accumulation} (\autoref{fig:examples}~(c)) for array-based containers \verb+std::string+.
This also bypasses the abstraction layer, as correct algorithm choice depends on container implementation internals (abstraction leakage).
The Haskell implementations also have limitations.
For example, the idiomatic combinator \verb+filter+ implementation (\autoref{fig:examples}~(d)) is compact and elegant, but it will construct a {\em deep copy} of the input with no structural sharing.\footnote{
For simplicity, we analyze Haskell under {\em strict} semantics.
The time/space complexities for Haskell under {\em lazy} semantics are equivalent to those of {\em strict} semantics in the worst case.}
This deep copy will occur even if the input contains no non-ASCII characters.
The recursive version (\autoref{fig:examples}~(e)) also suffers from the same problem, as well as exposing the data structure implementation (i.e., a list of \verb+Char+).
The final implementation (\autoref{fig:examples}~(f)) uses \verb+Seq+ from \verb+Data.Sequence+---a persistent tree-based sequence implementation, possibly allowing for structural sharing.
However, the sequence API lacks an iterator abstraction, and \autoref{fig:examples}~(f) circumvents this limitation by using explicit integer {\em indices}.
However, aside from the additional verbosity, deletion-by-index is an $O(\log N)$ operation, and thus the overall algorithm is a sub-optimal $O(N . \log N)$.
Finally, we note that existing \verb_C++_ persistent data structure libraries 
only expose read-only {\em const} iterators, and thus does not provide the iterator interface necessary to support the in-place updates required by \autoref{fig:examples}~(b).

In contrast, we consider the \tool version of \verb+filterAscii+ implemented using {\em persistent iterators} in \autoref{fig:examples}~(a).
This version uses familiar \verb_C++_-style iterator programming (similar to \autoref{fig:examples}~(b)), and achieves all four properties outlined above, including: the algorithmic/safety benefits of persistent data structures, an optimal $O(N)$ complexity, and abstracts the underlying container representation.
The main difference is that the iterator \verb+i+ implements {\em value} rather than reference semantics, meaning that the operation \verb_i.erase()_ erases the current character element independently of the originating container \verb+s+.
Furthermore, due to the lack of mutable reference semantics, the iterator-local copy must be explicitly extracted \verb+i.value()+.
The original container \verb+s+ and all copies are {\em unmodified} by the loop.

\begin{table}[t]
{\footnotesize
\centering
\renewcommand{\arraystretch}{1.2}
\setlength{\tabcolsep}{3pt}
\begin{tabular}{lccccccc}
\toprule
\textbf{Version} &
\textbf{Language} &
\textbf{API} &
\textbf{Complexity} &
\textbf{Sharing?} &
\textbf{Persistent?} &
\textbf{Abstract?} &
\textbf{Idiomatic?} \\
\midrule

\autoref{fig:examples}~\textcolor{red!60!black}{(b)} & \texttt{C++} & STL \texttt{std::list<char>} &
\textcolor{green!50!black}{$O(N)$} &
\cmark &
\xmark &
\nmark &
\xmark \\

\autoref{fig:examples}~\textcolor{red!60!black}{(b)}$^\dagger$ & \texttt{C++} & STL \texttt{std::string} &
\textcolor{red!50!black}{$O(N^2)$} &
\xmark &
\xmark &
\nmark &
\xmark \\

\autoref{fig:examples}~\textcolor{red!60!black}{(c)} & \texttt{C++} & STL \texttt{std::string} &
\textcolor{green!50!black}{$O(N)$} &
\xmark &
\xmark &
\nmark &
\cmark \\

\autoref{fig:examples}~\textcolor{red!60!black}{(c)}$^\ddagger$ & \texttt{C++} & \texttt{immer::vector<char>} &
\textcolor{green!50!black}{$O(N)$} &
\xmark &
\cmark &
\cmark &
\cmark \\

\autoref{fig:examples}~\textcolor{red!60!black}{(d)} & Haskell & \texttt{String} &
\textcolor{green!50!black}{$O(N)$} &
\xmark &
\cmark &
\cmark &
\cmark \\

\autoref{fig:examples}~\textcolor{red!60!black}{(e)} & Haskell & \texttt{String} (recursion) &
\textcolor{green!50!black}{$O(N)$} &
\xmark &
\cmark &
\xmark &
\cmark \\

\autoref{fig:examples}~\textcolor{red!60!black}{(f)} & Haskell & \texttt{Seq Char} &
\textcolor{red!50!black}{$O(N . \log N)$} &
\cmark &
\cmark &
\cmark &
\xmark \\

\midrule
\autoref{fig:examples}~\textcolor{red!60!black}{(a)} & \texttt{C++} & \tool \texttt{string} &
\textcolor{green!50!black}{$O(N)$} &
\cmark &
\cmark &
\cmark &
\cmark \\

\bottomrule
\end{tabular}
}
\vspace{1em}
\caption{Comparison of \texttt{filterAscii} implementations across different languages and paradigms.
We compare the worst-case algorithmic \emph{complexity}, whether the result \emph{shares} structure with the input, whether the underlying data structure is \emph{persistent}, whether the underlying data structure is \emph{abstracted}, and whether the implementation is considered \emph{idiomatic}.
Here, ($\dagger$) represents a version of \autoref{fig:examples}~(b) using \texttt{std::string} (with $O(N)$ erasure), ($\ddagger$) represents a version of \autoref{fig:examples}~(c) using Immer~\cite{puente17immer} (necessary since Immer only exposes const-iterators), and ($\nmark$) indicates the  data structure is partly revealed by the algorithmic complexity (leaky abstraction).
Only the \tool implementation with persistent iterators achieves sharing, persistence, and abstraction with optimal $O(N)$ algorithmic complexity.}
\label{tab:filterAscii}
\end{table}

\subsection{Contributions}

The \verb+filterAscii+ program is just one example, but linear scans of containers/data structures are a common programming pattern.
Yet existing programming paradigms either lack persistence (imperative) or are either sub-optimal in algorithmic or memory complexity (functional).
Persistent iterators combine key aspects of both approaches: iterators as a simple abstraction, persistence, and optimal time ($O(N)$ linear scan) and memory (structural sharing).
Persistent iterators thus retain the programming idiom of \verb_C++_ iteration, but not the reference semantics of the STL.
Developers can write new code (or adapt existing algorithms) under a persistent, value-based paradigm while keeping a familiar iterator-style API.
In summary, the main contributions of this paper are:
\begin{itemize}[leftmargin=*]
\item We introduce the concept of {\em persistent iterators}---bridging the gap between traditional \verb_C++_ iterator-based programming with the safety and power of persistent data structures.
Unlike traditional mutable iterators that implement {\em reference semantics}, persistent iterators implement {\em value semantics} where each iterator object works on its own local copy of the underlying data structure---eliminating the risk of iterator invalidation bugs.
And unlike const-iterators, persistent iterators allow for the (local copy) of the underlying data structure to be updated (assignment, insertion, erasure)---supporting a more traditional \verb_C++_ iterator programming paradigm where containers can be modified via iterators.
\item We show that persistent versions of common containers (\emph{vectors}, \emph{sets}, \emph{maps}, \emph{strings}, etc.) can be implemented using a single underlying data structure---the {\em finger tree}~\cite{HINZ2006}---achieving a comparable algorithmic complexity with the \verb_C++_ {\em Standard Template Library} (STL) operations.
We show that persistent iterators can be implemented as optimized {\em zippers}~\cite{HUET1997} into persistent finger trees, and can be used to implement all of the standard \verb_C++_ STL iterator operations (increment, decrement, dereference, etc.)---again with comparable algorithmic complexity.
Our implementation is {\em uniform}---i.e., all containers use the same underlying data structure, preventing asymptotic surprises or the need for container-specific special cases (see \autoref{fig:examples}~(b) and (c)).
To our knowledge, we are also the first implementation of zippers over finger trees.
\item We implement our approach in the form of \tool---a \verb_C++_ container library that supports persistent versions of common containers.
We evaluate \tool against various popular \verb_C++_ libraries, including the STL~\cite{stepanov1995stl}, Immer~\cite{puente17immer}, Abseil~\cite{Abseil}, and Folly~\cite{Folly}.
We show that \tool improves the safety in the usage of the containers with performance comparable to other persistent libraries.
\end{itemize}
This paper's core contribution is a semantic reconciliation: we show that iterator-based imperative programming can be reconciled with full value semantics over persistent data structures, without sacrificing expressiveness.

\myparagraph{Non-goals}
\tool is not intended as a drop-in replacement for STL containers in performance-critical code.
Persistent data structures incur inherent constant-factor overheads and cannot match contiguous array-based implementations.
Our goal is instead to provide an STL-like alternative for persistence-oriented use cases, where safety and value semantics are primary.

\section{Background}

\subsection{Iterators}

The {\em iterator abstraction}~\cite{Gamma95, stepanov1995stl} is widely used to provide a uniform way to access an element of a data structure in a sequential manner.
Each {\em iterator} refers to a concrete element inside the structure (or a one-past-the-end special case), and iterators provide uniform operations that can either traverse or modify the parent collection.
Modern {\em Standard Template Library} (STL) iterators can be advanced or receded to refer to different elements, and support element retrieval, replacement, or erasure at the specified position.
Iterators are highly expressive: combinations of these simple operations can be used to implement many common algorithms.
Furthermore, iterators are also abstract, allowing for element traversal, retrieval, and modification in a uniform way that does not directly expose the underlying container representation.

\subsubsection{STL Iterators in \texttt{C++}}

Iterators form the backbone of containers and algorithms in the \verb_C++_ STL.
Most STL algorithms express their inputs and outputs exclusively in terms of iterators or iterator pairs, thus enabling their reuse against different container types.
To respect this abstraction, each standard container also provides \texttt{begin()} and \texttt{end()} member functions that return iterators referring to the first and one-past-the-end element, respectively.
In \verb_C++_, iterators themselves are modeled as an abstraction of a \verb+C+-style pointer, thus having similar usage syntax.
For example, \verb_++it_ advances the iterator to the next element in the container (conceptually similar to {\em pointer arithmetic}), and \verb_*it_ fetches the current element referred to by the iterator (conceptually similar to {\em pointer dereference}).
Furthermore, \verb_C++_ is among the few languages that split iterator advancement, the testing of validity, and dereferencing into three separate operations.
While this model provides maximum flexibility, it also increases the complexity of iterator implementations.

STL Iterators themselves are typically implemented as lightweight copy-constructable objects, only containing a pointer that directly references a specific element in the originating container.
Thus, iterators effectively implement a form of {\em reference semantics}, where the originating container can be accessed or even directly mutated via an iterator.
However, the complexity of the API and reference semantics can lead to subtle errors when programming with iterators.

\begin{figure}
\begin{minipage}{0.47\linewidth}
\centering
\lstset{style=CStyle, backgroundcolor=\color{red!8}}
\begin{lstlisting}
void filterAscii(std::list<char32_t> &s) {
  for (auto i = s.begin(); i != s.end(); )
    if (*i >= 0x7F) s.erase(i);  // BUG!
    else            ++i;
}
\end{lstlisting}
\end{minipage}
\hfill
  \begin{minipage}[c]{0.5\textwidth}
    \vspace{-1em}
    \caption{
    Example of a classic {\em iterator invalidation} bug variant of \autoref{fig:examples}~(b).
    Here, the result of \texttt{s.erase(i)} has not been reassigned back to \texttt{i}.
    } \label{fig:invalid}
  \end{minipage}
\end{figure}

\subsection{Limitations of STL Iterators}\label{sec:iterator_misuse}

While iterators are a powerful abstraction, they also have some well-known limitations that are sometimes hard to mitigate.
We present below several common misuses of STL iterators.

\myparagraph{Iterator Invalidation}
As \verb_C++_ iterators are designed as a conceptual extension of the pointer model, they also suffer from the same pitfalls.
When the underlying memory location managed by the parent container is mutated or de-/re-allocated, the iterators pointing to this location may be {\em invalidated}.
In a low-level programming language such as \verb_C++_, iterator invalidation can be a source of vulnerability, such as {\em use-after-free} (UaF) errors.
A common example of iterator invalidation is illustrated in \autoref{fig:invalid}.
Here, the \verb+s.erase(i)+ operation will invalidate iterator \verb+i+, and the correct idiom is to reassign the result of the erasure (see \autoref{fig:examples}~(b)).

A related problem is the possibility of iteration occurring concurrently with modifications to the container itself ({\em concurrent modification}).
For example, if elements are added to a container during iteration, the previously obtained iterator may or may not traverse the newly added elements, depending on the underlying semantics of the container.
Each container and operation has its own iterator invalidation rules.
For example, the rules for \verb+vector::shrink_to_fit+ and \verb+vector::erase+ are very different: with the former (almost) always invalidating all iterators (unless no change in capacity), and the latter invalidating iterators only {\em after} the erased element.
Many languages, such as Java and C\#, attempt to detect invalidation and concurrent modification {\em dynamically}, and throw an error in a {\em fail-fast} manner (e.g., Java's \texttt{ConcurrentModificationException}~\cite{java_collections_framework}).
Rust detects potential invalidation at {\em compile-time} using static borrow and lifetime checking.
Python is memory safe by design,
but may exhibit unintuitive behaviors if a container is modified during iteration.

\myparagraph{Data Structure Invariants}
In low-level languages such as \verb_C++_, mutation via iterators can also lead to subtle issues, since elements may be modified in place.
While standard containers restrict certain modifications to preserve invariants (e.g., keys in associative containers), indirect mutation through shared or aliased state can still violate logical invariants in practice.
As a recent example, modification through \texttt{filter\_view}'s iterator can cause elements satisfying the given predicate to be skipped~\cite{Nicolai24}.
The lack of deep-\texttt{const}ness in the language allows mutation through ostensibly immutable references~\cite{Eyolfson16}.

\myparagraph{Uniformity and Leaky Abstractions}
Although iterators provide a uniform abstraction in principle, container implementation details can still leak via asymptotic complexities.
For example, the implementations of \autoref{fig:examples}~(b) and (c) are very different due to differences in the underlying container implementation.
While porting \autoref{fig:examples}~(b) to \verb+std::string+ is possible, the erasure operation becomes $O(N)$, making the overall algorithm $O(N^2)$ in the worst case.

\subsection{Persistent Data Structures}\label{sec:persistent_data_structures}

In conventional imperative programming, containers (e.g., \verb_C++_ STL) are {\em mutable}: operations like \verb+insert+, \verb+erase+, or assignment modify the container in place.
Mutation is efficient and familiar, but it tightly couples all references to a single, ever-changing state.
Any update can invalidate iterators, aliases, or cached computations, and reasoning about program behavior requires tracking these side effects across potentially complex aliasing relationships.

{\em Persistent data structures} take a fundamentally different approach.
A data structure is {\em persistent} if every update produces a new version of the structure while leaving the original version intact.
The update does not overwrite (mutate) or destroy existing nodes; rather, it allocates {\em new} nodes only along the path that the update changes, and all unmodified portions of the structure are {\em structurally shared} between the old and new versions.
For example, inserting an element into a persistent balanced tree allocates new nodes only along the search path, while all subtrees not affected by the insertion will be {\em shared}.
The original and updated versions can coexist safely---providing logical immutability without ``deep'' copying of the entire structure.
Persistent data structures, therefore, offer a number of semantic and practical advantages.
Firstly, they eliminate side effects: an update cannot modify any existing version, removing an entire class of aliasing bugs.
Secondly, ``copying'' a persistent container becomes a cheap $O(1)$ operation,
making {\em value semantics} practical and efficient: assigning a container simply refers to the same underlying structure until one of the copies is modified. Because every version is logically independent, persistent containers are inherently thread-safe for concurrent reads and allow fine-grained rollback, undo, and snapshot semantics.

\myparagraph{Memory Management}
Because each node may be pointed to by more than one parent, memory {\em reclamation} must account for this sharing.
{\em Garbage collection} (GC) naturally supports persistence by reclaiming unreachable nodes automatically.
In manual-memory settings such as \verb_C++_, {\em reference counting} provides a practical alternative: since our persistent data structures are acyclic by construction (nodes are immutable and references flow only from parents to children), simple reference counting suffices to reclaim storage deterministically.

\myparagraph{Limitations of Persistent Data Structures}
The benefits of persistent data structures come with some costs and design trade-offs.
To achieve efficient updates, persistent structures usually rely on tree-like representations that support structural sharing.
Array-based structures such as \verb_std::vector_ or \verb_std::string_ cannot simply copy a single contiguous buffer on every modification without losing asymptotic efficiency.
Instead, practical persistent vectors and strings (as in Clojure or immer~\cite{puente17immer}) are implemented as trees of fixed-size chunks.
Even then, updates allocate new nodes, producing transient garbage when older versions are no longer referenced.
This can make persistent versions of mutable containers inherently slower, especially for array-based containers.
We refer to this as the ``persistence tax'', and it is generally the main trade-off when using persistent data structures.
That said, we aim to minimize the impact as much as possible.

\subsection{Value Semantics}
In programming languages, \emph{value semantics} refers to a model in which variables hold values rather than references to shared mutable state. 
When a value is assigned, copied, or passed to a function, the new variable conceptually receives its own independent copy. 
This is the semantics used by primitive types in \verb_C++_.
For example, consider the two program fragments:

{\small
\definecolor{darkgreen}{RGB}{0,100,0}
\definecolor{darkred}{RGB}{100,0,0}
\begin{Verbatim}[commandchars=\\\{\}]
            \textcolor{darkgreen}{int} a = \textcolor{darkred}{10}; \textcolor{darkgreen}{int} b = a; b++;        \textcolor{darkgreen}{int} x = \textcolor{darkred}{10}; \textcolor{darkgreen}{int &}y = x; y++;
\end{Verbatim}
}

\noindent
The first uses {\em value semantics}, meaning that the update \verb_b++_ does {\em not} affect the value of \verb+a+.
By contrast, the second uses \emph{reference semantics} that associate variables with shared objects in memory, so the update \verb_y++_ alters the observable state of \verb+x+---i.e., a {\em side effect}. 
Most standard STL containers (\verb|std::vector|, \verb|std::map|, etc.) follow reference semantics for their iterators and internal storage, which can lead to subtle invalidation errors when the underlying container is mutated.

\begin{figure}[t]
\centering

\lstdefinestyle{cppsmall}{
  language=C++,
  basicstyle=\ttfamily\scriptsize,
  keywordstyle=\color{blue},
  commentstyle=\color{green!50!black},
  stringstyle=\color{orange!60!black},
  showstringspaces=false,
  columns=fullflexible,
  numberstyle=\tiny\color{gray},
  numbersep=6pt,
  keepspaces=true,
  frame=single,
  framerule=0.3pt,
  xleftmargin=2pt,
  xrightmargin=2pt,
  backgroundcolor=\color{gray!3},
  morekeywords={string,vector,size_t,iterator},
  literate=
 {0}{{{\color{red!70!black}0}}}{1}
 {1}{{{\color{red!70!black}1}}}{1}
 {2}{{{\color{red!70!black}2}}}{1}
 {3}{{{\color{red!70!black}3}}}{1}
 {4}{{{\color{red!70!black}4}}}{1}
 {5}{{{\color{red!70!black}5}}}{1}
 {6}{{{\color{red!70!black}6}}}{1}
 {7}{{{\color{red!70!black}7}}}{1}
 {8}{{{\color{red!70!black}8}}}{1}
 {9}{{{\color{red!70!black}9}}}{1},
}

\lstdefinestyle{hssmall}{
  language=Haskell,
  basicstyle=\ttfamily\scriptsize,
  keywordstyle=\color{blue},
  commentstyle=\color{green!50!black},
  stringstyle=\color{orange!60!black},
  showstringspaces=false,
  columns=fullflexible,
  numberstyle=\tiny\color{gray},
  numbersep=6pt,
  keepspaces=true,
  frame=single,
  framerule=0.3pt,
  xleftmargin=2pt,
  xrightmargin=2pt,
  backgroundcolor=\color{gray!3},
}

\vspace{0.5em}
\begin{minipage}{0.48\linewidth}
\begin{lstlisting}[style=hssmall, numbers=left, escapeinside={(*@}{@*)}]
filterReachable :: [Stmt] -> [Stmt]
filterReachable =
 reverse . snd . foldl step (True, [])
 where
  step (reach, acc) stmt =
   let reach1 = if isLabel stmt then True else reach
       acc'   = if reach1 then stmt : acc else acc
       reach2 = if isGoto stmt then False else reach1
   in (reach2, acc')
(*@ @*)
\end{lstlisting}
\vspace{-0.5em}
\centerline{\small (a) Haskell combinator implementation}
\end{minipage}
\hspace{0.2em}
\begin{minipage}{0.44\linewidth}
\begin{lstlisting}[style=cppsmall]
 bool reach = true;
 auto it = stmts.begin();
 for (; it != stmts.end(); ) {
  if (isLabel(*it)) reach = true;
  if (reach) {
   if (isGoto(*it)) reach = false;
   ++it;
  } else it.erase();
 }
 stmts = it.value();
\end{lstlisting}
\vspace{-0.5em}
\centerline{\small (b) Persistent iterator implementation}
\end{minipage}

\vspace{0.5em}
\begin{minipage}{0.48\linewidth}
\begin{lstlisting}[style=cppsmall, numbers=left,
linebackgroundcolor={
                     \color{gray!3}
                     \ifnum\value{lstnumber}=2\color{yellow!20}\fi
                     \ifnum\value{lstnumber}=6\color{red!15}\fi
                     \ifnum\value{lstnumber}=11\color{yellow!20}\fi
                     \ifnum\value{lstnumber}=12\color{yellow!20}\fi
                     \ifnum\value{lstnumber}=18\color{yellow!20}\fi
                     \ifnum\value{lstnumber}=28\color{red!15}\fi
                     \ifnum\value{lstnumber}=29\color{yellow!20}\fi
                     \ifnum\value{lstnumber}=34\color{red!15}\fi
                     \ifnum\value{lstnumber}=35\color{yellow!20}\fi
                   }]
struct Editor {
  struct Cursor { string buf; size_t pos = 0; };
  Cursor cursor;
  struct { vector<Cursor> undo, redo; } history;
  void commit() {
   history.undo.push_back(cursor);
   history.redo.clear();
  }
  void insert(char c) {
   commit();
   cursor.buf.insert(cursor.pos, 1, c);
   cursor.pos++;
  }
  void backspace() {
   if (cursor.pos == 0) return;
   commit();
   cursor.pos--;
   cursor.buf.erase(cursor.pos);
  }
  void left() {
   if (cursor.pos > 0) cursor.pos--;
  }
  void right() {
   if (cursor.pos < cursor.buf.size()) cursor.pos++;
  }
  void undo() {
   if (history.undo.empty()) return;
   history.redo.push_back(cursor);
   cursor = std::move(history.undo.back());
   history.undo.pop_back();
  }
  void redo() {
    if (history.redo.empty()) return;
    history.undo.push_back(cursor);
    cursor = std::move(history.redo.back());
    history.redo.pop_back();
  }
};
\end{lstlisting}
\vspace{-0.5em}
\centerline{\small (c) STL editor implementation}
\end{minipage}
\hspace{0.2em}
\begin{minipage}{0.44\linewidth}
\begin{lstlisting}[style=cppsmall, linebackgroundcolor={
                     \color{gray!3}
                     \ifnum\value{lstnumber}=2\color{yellow!20}\fi
                     \ifnum\value{lstnumber}=6\color{green!15}\fi
                     \ifnum\value{lstnumber}=11\color{yellow!20}\fi
                     \ifnum\value{lstnumber}=12\color{yellow!20}\fi
                     \ifnum\value{lstnumber}=18\color{yellow!20}\fi
                     \ifnum\value{lstnumber}=28\color{green!15}\fi
                     \ifnum\value{lstnumber}=29\color{yellow!20}\fi
                     \ifnum\value{lstnumber}=34\color{green!15}\fi
                     \ifnum\value{lstnumber}=35\color{yellow!20}\fi
                   }]
struct Editor {
  using Cursor = string::iterator;
  Cursor cursor;
  struct { vector<Cursor> undo, redo; } history;
  void commit() {
    history.undo.push_back(cursor);
    history.redo.clear();
  }
  void insert(char c) {
    commit();
    cursor.insert(c);
    
  }
  void backspace() {
    if (cursor.pos() == 0) return;
    commit();
    cursor--;
    cursor.erase();
  }
  void left() {
    if (cursor.pos() > 0) cursor--;
  }
  void right() {
    if (cursor != string::end()) cursor++;
  }
  void undo() {
    if (history.undo.empty()) return;
    history.redo.push_back(cursor);
    cursor = history.undo.back();
    history.undo.pop_back();
  }
  void redo() {
    if (history.redo.empty()) return;
    history.undo.push_back(cursor);
    cursor = history.redo.back();
    history.redo.pop_back();
  }
};
\end{lstlisting}
\vspace{-0.5em}
\centerline{\small (d) Persistent iterator editor implementation}
\end{minipage}

\caption{Sample programs that illustrate the advantages of persistent iterators,
including stateful filtering and a minimal text editor.
We compare Haskell/STL container implementations (left) and persistent iterator implementations (right).
Here, \colorbox{green!15}{green}${=}$ $O(1)$ shallow copy, \colorbox{red!15}{red}${=}$ $O(N)$ deep copy, and \colorbox{yellow!20}{yellow}${=}$major differences.
Note that some lines are identical syntactically but differ in complexity.
}
\label{fig:editor}
\end{figure}

Value semantics are desirable because they simplify reasoning about program behavior. 
Updates have no side effects beyond the scope of the variable being modified, supporting safer and more modular code. 
This model closely parallels the principles of {\em functional programming}, where functions operate on immutable values and return new ones. 
In practice, however, deep copying of large data structures is inefficient, which has historically limited the use of value semantics to small or immutable types.
However, persistent data structures overcome this limitation through structural sharing. 
It is also important to clarify what value semantics \emph{do not} mean.
Value semantics do not require variables themselves to be immutable or prohibit reassignment (as would be required for full {\em referential transparency}).
For example, consider the program fragment:

{\small
\definecolor{darkgreen}{RGB}{0,100,0}
\definecolor{darkred}{RGB}{100,0,0}
\begin{Verbatim}[commandchars=\\\{\}]
                    \textcolor{darkgreen}{vector} a(\{\textcolor{darkred}{2},\textcolor{darkred}{3},\textcolor{darkred}{5}\}); \textcolor{darkgreen}{vector} b = a; b.push_back(\textcolor{darkred}{7});
\end{Verbatim}
}

\noindent
Programmers can freely write statements such as \verb|b.push_back(7)|, which only changes the value referenced by \verb|b| and not other values, such as those referenced by \verb|a|.
Value semantics ensure that each variable refers to an independent logical value, even if the underlying storage is shared. 

\subsection{Additional Examples and Programming Patterns}

Persistent iterators sometimes enable simpler or more efficient formulations over their counterparts.
In addition to the motivating example in \autoref{sec:example}, we present two small examples that highlight situations where persistent iterators can simplify common programming patterns.

\myparagraph{Pattern 1: Stateful filtering}
Given a sequence of \verb+C+-style statements (including labels and gotos), \autoref{fig:editor}~(a) and (b) remove any statement that is unreachable due to prior control flow.
This pattern can be expressed in Haskell using {\em recursion}, but this relies on an underlying list representation.
The pattern falls outside the scope of common Haskell {\em combinators} (e.g., \verb+map+, \verb+filter+, etc.), but can be expressed by threading a {\em reachability flag} through a general {\em left fold} (\verb+foldl+) followed by an explicit \verb+reverse+ pass to restore ordering (\autoref{fig:editor}~(a)).
In contrast, the \tool version (\autoref{fig:editor}~(b)) is expressed as a single-pass iterator loop, where the reachability state is maintained explicitly, and all updates occur under value semantics.
This formulation is both natural and idiomatic, and remains fully abstract: the same code applies uniformly across container types without exposing the underlying data-structure.
Finally, the \tool version exploits {\em structural sharing} between the input and output, whereas the Haskell version always constructs a new list.

\myparagraph{Pattern 2: Snapshot-based state management}
\autoref{fig:editor}~(c) and~(d) show a minimal text editor supporting {\em insert}, {\em backspace}, {\em cursor movement}, and {\em undo}/{\em redo}.
The editor state is represented by a cursor and history stacks storing prior states.

In the STL implementation (\autoref{fig:editor}~(c)), storing editor state requires an $O(N)$ deep copy of the entire string buffer.
As such, a simple snapshot-based design is not practical, and real text editors typically use more complex designs (e.g., {\em command logs}).
With \autoref{fig:editor}~(d), the cursor is naturally represented as a {\em persistent iterator} encoding both the buffer and position.
Editor operations map directly to iterator operations (\verb_++_, \verb_--_, \verb_insert_, \verb_erase_).
This enables a design where editor state can be captured and restored directly as {\em values}, without auxiliary data structures or restructuring of the algorithm.
In contrast to traditional containers, persistent iterators make this design both practical and idiomatic: the cursor is a value that encodes both position and structure, and undo/redo reduces to storing snapshots in $O(1)$ time via structural sharing.
This pattern generalizes to other applications requiring undo/redo, speculative execution, or versioned state.

\section{Design of Persistent Containers}

Our persistent container design is built on top of {\em finger trees}~\cite{HINZ2006}---a general-purpose, persistent data structure that provides asymptotic performance comparable to mutable containers such as those in the \verb_C++_ STL.
For example, and uniquely for finger trees, adding or removing elements from the back (or front) of a tree is an amortized $O(1)$ operation.
Similarly, in imperative settings, \verb|std::vector::push_back| or \verb+pop_back+ is also an amortized $O(1)$ operation, by storing elements in a contiguous buffer that occasionally resizes and copies.
Thus, and already for this example, finger trees are a good fit for a persistent vector implementation: both can be used to represent sequences of elements with a similar algorithmic complexity.
 
\subsection{Finger Trees}

A {\em finger tree}~\cite{HINZ2006} is a specially balanced tree optimized for sequences and fast access near its ends.
Conceptually, a finger tree consists of a list of recursive {\em spine nodes}, each of which emits two {\em finger nodes} (a left and a right) representing the sequence {\em prefix} and {\em suffix} of the child spine node.
Each finger node stores typically 1-4 {\em digits}, where each digit is a balanced 2-3 tree containing the actual elements of the sequence.
The key to finger trees is the depth of each digit, with the top-most spine node comprising digits of depth 0 (i.e., individual elements), the second top-most comprising digits of depth 1, etc.
Thus, accessing the back (or front) of a finger tree is therefore an $O(1)$ operation, as only the root spine node and rightmost (or leftmost) digit need be accessed.
Furthermore, the overall finger tree structure is balanced and maintains logarithmic depth relative to the total number of elements.
Thus, top-level fingers provide constant-time access to both ends, while the recursive spine ensures logarithmic access to the interior.

Inserting an element (at either end) modifies only the shallow digits of the corresponding finger.
When a digit overflows, a small constant number of nodes (typically 2) are recursively promoted down the spine, maintaining balance.
As a result, inserting an element requires only touching $O(1)$ nodes on average.
Similarly, deleting an element (at either end) is also an amortized $O(1)$ operation.
In addition, finger trees support appending entire sequences, splitting, and accessing/insertion/deletion of elements at random positions, all in $O(\log N)$ time.
And since finger trees are trees, updates can be made {\em persistently} by re-allocating only the nodes along the modified path while reusing all unaffected subtrees for structural sharing.
The asymptotic time complexities of finger trees closely resemble their mutable STL counterparts, but with persistence guarantees.

\begin{figure}
\begin{minipage}{0.48\textwidth}
\fbox{\includegraphics[scale=0.9]{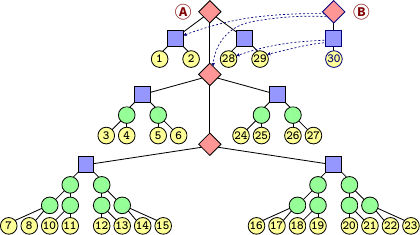}}
\end{minipage}
\hfill
\begin{minipage}{0.51\textwidth}
    \caption{Example finger trees representing the sequences \textcolor{red!50!black}{\circled{A}}${=}\langle 1, 2, ..., 8, 10, ... 29 \rangle$ and \textcolor{red!50!black}{\circled{B}}${=}\langle 1, 2, ..., 8, 10, ... 30 \rangle$.
Here, {\em spine} nodes are represented by red diamonds, {\em finger nodes} by blue boxes, 2-3 tree nodes as green circles, and {\em element} nodes by yellow circles with the corresponding value.
Furthermore, tree \textcolor{red!50!black}{\circled{B}} is created via a \texttt{push\_back(30)} operation on tree \textcolor{red!50!black}{\circled{A}}, and benefits from structural sharing.
Here, dashed edges represent references from \textcolor{red!50!black}{\circled{B}} to the original \textcolor{red!50!black}{\circled{A}} tree nodes.
\label{fig:finger}}
\end{minipage}
\end{figure}

\myparagraph{Example}
Example finger trees are illustrated in \autoref{fig:finger}.
Here, tree \textcolor{red!50!black}{\circled{A}} illustrates the unique finger tree structure, with a central spine of nodes (red diamonds), each with two fingers (blue boxes), with some number of 2-3 tree digits (green/yellow circles) of increasing depth.
This example assumes a branching factor of 2, but in general fingers may have 1-4 digits, and tree nodes may have 2-3 children (i.e., 2-3 trees).
Tree \textcolor{red!50!black}{\circled{B}} is constructed from \textcolor{red!50!black}{\circled{A}} using a \verb+push_back(30)+ operation.
Since this updates the end of the structure, only a limited number of nodes (namely, 3) need to be copied, and a significant portion of \textcolor{red!50!black}{\circled{A}}'s structure is shared by \textcolor{red!50!black}{\circled{B}}.
The finger tree structural invariant and operations are complex, so we omit many details for brevity (see the original paper~\cite{HINZ2006} for details).

\subsubsection{Persistent Vectors as Finger Trees}

Persistent vectors are implemented directly as finger trees.
Each \verb+push_back+, \verb+pop_back+, or \verb+back+ operation only touches the right-hand finger, providing amortized $O(1)$ performance analogous to \verb+std::vector+.
The difference lies in semantics: rather than mutating the container in place, \verb+push_back+ allocates a few new nodes to represent the updated path and returns a new persistent vector, meaning that any previous version will remain valid.

\subsubsection{Finger Trees as a Unifying Abstraction (Persistent Containers as Finger Trees)}

A major insight of our design is that the finger tree serves as a single unifying abstraction for {\em all} common container types---not only {\em vectors} but also {\em (multi)sets}, {\em (multi)maps}, and {\em strings}.
These conventional containers only fundamentally differ in the invariants that are maintained over the underlying sequence.
This gives rise to a natural type hierarchy:
\begin{itemize}[leftmargin=*]
\item[-] A {\em vector} is a sequence;
\item[-] A {\em multiset} is a sorted vector;
\item[-] A {\em set} is a multiset with no duplicate elements;
\item[-] A {\em multimap} is a multiset of key-value pairs ordered by key;
\item[-] A {\em map} is a multimap with unique keys; and
\item[-] A {\em string} is a vector of characters under the UTF-8 encoding.
\end{itemize}
Essentially, {\em all} common container types are special cases of sequences, and sequences have a universal persistent implementation: the finger tree.
In our implementation, each container is realized as a finger tree equipped with a specialized {\em monoid}~\cite{HINZ2006} summarizing subtree information.
For {\em (multi)sets} and {\em (multi)maps}, the monoid is the {\em minimum element} of the left subtree, and for {\em strings}, the monoid is the total Unicode length under the UTF-8 encoding (rather than the total single-byte characters).
All containers, therefore, share the same structural backbone and persistence mechanism, and only their monoid semantics differ.
Since vectors already use a tree-based representation, accessing/insertion/deletion of elements in sets and maps still uses a tree-based search in $O(\log N)$ time---the same as the STL counterparts.
This uniform design yields three significant benefits:
\begin{itemize}[leftmargin=*]
\item {\em Zero-cost upcasts}: containers can be safely upcast without conversion overhead.
For example, viewing a set as a vector requires no copying since the underlying data representation is identical.
Similarly, UTF8-strings can be viewed as vectors of single-byte characters.
\item {\em Semantic uniformity}: algorithms can be written generically over a common finger-tree implementation, avoiding multiple diverging container-specific implementations.
\item {\em Implementation simplicity}: the entire library is built atop a single core data structure, greatly reducing duplication and the potential for asymptotic ``surprises'' across container types.
\end{itemize}
For example, the algorithmic choice in \autoref{fig:examples}~(b) and (c) depends on the implementation details of the underlying container type, which is a form of abstraction leakage.
Under our design, the problem would not exist, since all containers use the same unified finger tree implementation.

\section{Design of Persistent Iterators}

Having established a persistent foundation for containers, we now turn to {\em persistent iterators}---our mechanism for traversing and modifying containers while retaining full persistence.
The goal is to provide a familiar, STL-style iterator interface without reintroducing the aliasing and invalidation hazards that arise from the default STL iterator {\em reference semantics}.

\subsection{Persistent Iterators as Zippers}

We implement persistent iterators using a variant of the {\em zipper}~\cite{HUET1997}, a pure functional data structure that represents a ``cursor'' or ``position'' into another persistent structure.
A zipper decomposes a tree (such as a finger tree) into a {\em reversed path} from a node (such as a leaf element) back to the tree's root, capturing the context necessary to reconstruct the full tree structure.
This representation enables direct $O(1)$ access to the {\em current element}, which is at the {\em head} of the reversed path, rather than $O(\log N)$ access from the root.
Navigation operations (\verb_++_, \verb_--_, etc.) are implemented by moving up the reversed path to the shared ancestor, and then down a neighboring branch.

\newcommand{\treenode}[0]{\mathit{tree{\text -}node}}
\newcommand{\childindex}[0]{\mathit{child{\text -}index}}
\newcommand{\dirtyflag}[0]{\mathit{dirty{\text -}flag}}
\newcommand{\parentnode}[0]{\mathit{parent{\text -}node}}

In our implementation, each zipper node encodes a tuple:
\begin{align*}
\langle \treenode, \childindex, \dirtyflag, \parentnode \rangle
\end{align*}
Here, $\treenode$ is the finger tree node in the path, $\childindex$ indicates which child is taken along the path, $\dirtyflag$ (explained below) indicates whether $\treenode$ has been modified from the parent, and $\parentnode$ is the zipper node containing the parent of $\treenode$ (or \emph{null} for the root node).
Unlike the original zipper design~\cite{HUET1997}, our variant minimizes allocations by storing direct references to existing finger tree nodes along with their child indices, rather than duplicating path information.
The top-level zipper head node corresponds to the current element, and its parent pointer recursively encodes the rest of the reversed path toward the root.

\begin{figure}
\begin{minipage}{0.32\textwidth}
\caption{Example finger tree \textcolor{red!50!black}{\circled{A}} with a zipper \textcolor{red!50!black}{\circled{Z}} pointing to element \texttt{8}, and a zipper \textcolor{red!50!black}{\circled{Y}} pointing to a newly {\em inserted} element \texttt{9}.
Note that zippers are also persistent structures with complex structural sharing relationships (even between zippers), as represented by the dashed lines.
\label{fig:zipper}}
\end{minipage}
$\quad$
\begin{minipage}{0.62\textwidth}
\fbox{\includegraphics[scale=0.9]{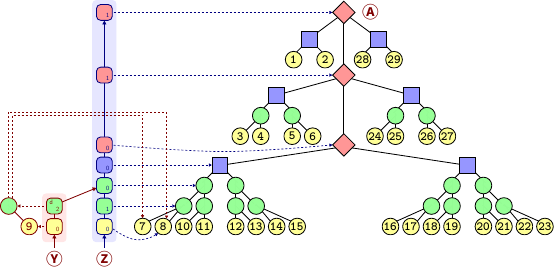}}
\end{minipage}
\end{figure}

\myparagraph{Example}
An example finger tree \textcolor{red!50!black}{\circled{A}} and zipper \textcolor{red!50!black}{\circled{Z}} is illustrated in \autoref{fig:zipper}.
The \textcolor{red!50!black}{\circled{Z}} zipper is a list-like structure starting from the bottom up.
The head of the zipper contains a reference to an {\em element node} of the finger tree, in this case, the node containing \verb+8+.
Thus, the element can be accessed in $O(1)$ time from the head node of the zipper, as opposed to $O(\log N)$ time from the tree root.
The remaining zipper nodes contain references to finger tree nodes and a corresponding $\childindex$ (represented as a sub-script), and thus are a complete encoding of a finger tree and (reversed) path to a specific element.

\subsubsection{Navigation}

Iterator movement corresponds to reconstructing or extending the reversed path.
To move from one position to another, the zipper ascends until it reaches the {\em lowest common ancestor} of the source and destination elements, and then descends along the new path.
For local operations such as increment \verb_++it_ or decrement \verb_--it_, this ancestor is typically close, yielding amortized $O(1)$ complexity.
Larger jumps, such as (\verb_it += N_), take $O(\log N)$ time in general.
Importantly, zippers---and therefore iterators---are themselves persistent.
Multiple iterators derived from the same container will share common path segments (i.e., {\em structural sharing}), and modifying or advancing one iterator does not invalidate or modify any other.

\subsubsection{Element Update, Insertion, and Deletion}

Persistent iterators support in-place-style updates, such as {\em assignment}, {\em insertion}, and {\em erasure}, without violating persistence.
To achieve this, we implement a lazy update mechanism based on the zipper's $\dirtyflag$.
When an element is replaced (assigned) to new value ($x$) via an iterator \verb+i+, a new iterator value \verb+j+ will be created by:
\begin{itemize}[leftmargin=*]
\item[-] Allocating a new finger tree leaf node $n$ with $x$ as the updated element value;
\item[-] Allocating a new zipper head node \verb+j+ that references the new element node; and
\item[-] Marking the new zipper head's $\dirtyflag$ as set.
\end{itemize}
In other words, the new iterator value \verb+j+ is defined as follows:
\begin{align*}
    \texttt{j} = \langle n,~\texttt{i}.\childindex,~\mathit{true},~\texttt{i}.\parentnode \rangle
\end{align*}
As such, assignment is an $O(1)$ operation.
The dirty flag indicates that the current subtree below the parent has been updated and must be reconstructed upon traversal.
When the iterator subsequently moves, the dirty flag is propagated upward, causing the affected finger tree nodes to be lazily rebuilt along the modified path until eventually reaching the root.
This design allows multiple element updates to be ``stacked'' efficiently, without requiring a full $O(\log N)$ path reconstruction for each update.

Element {\em insertion} and {\em deletion} (a.k.a. {\em erasure}) are handled similarly.
However, since these operations may affect local balance, the zipper will unwind to the nearest ancestor such that balance can be restored.
The unwinding depth is amortized $O(1)$ steps.
Once the finger tree invariant is restored, the zipper node is marked as dirty, and the operation completes.
As a result, persistent iterator assignments, insertions, and deletions can also be ``stacked'', and each operation runs in (amortized) $O(1)$ time for localized updates while preserving persistence and structural sharing.

\myparagraph{Example}
Another zipper \textcolor{red!50!black}{\circled{Y}} is also illustrated in \autoref{fig:zipper}.
Here, the missing element \verb+9+ has been {\em inserted} into the sequence through the original zipper \textcolor{red!50!black}{\circled{Z}}.
This involves unwinding the original zipper to the nearest ancestor such that a new balanced subtree can be constructed.
In this example, we need to unwind to depth 1 and convert the corresponding 2-node into a 3-node which holds the additional \verb+9+ element.
The modified finger tree is {\em not} reconstructed in its entirety.
Rather, the $\dirtyflag$ ($d$) is merely marked on the corresponding zipper node, and the modified finger tree will be reconstructed {\em lazily}.

Crucially, the new zipper \textcolor{red!50!black}{\circled{Y}} is still a {\em persistent} data structure, and structurally shares its tail with the original zipper \textcolor{red!50!black}{\circled{Z}}.
A complex web of structural sharing references is created, as illustrated in \autoref{fig:zipper}, and deciphering this web is left as an exercise for the reader.
Nevertheless, the original finger tree \textcolor{red!50!black}{\circled{A}} and the original zipper \textcolor{red!50!black}{\circled{Z}} remain {\em valid} and have not been mutated---i.e., this yields side-effect-free value semantics.

\subsubsection{Value Extraction}

Because persistent iterators possess {\em value semantics} rather than {\em reference semantics}, an iterator may represent a locally modified view of the container that differs from the original version.
To materialize this view as a first-class container value, we provide the \verb+it.value()+ operation that returns the iterator's current logical container---a persistent version that reflects all updates made through the iterator.
The implementation of \verb+value+ is conceptually simple:
starting from the iterator's current zipper node, the operation traverses the reversed path back to the root of the finger tree, an $O(\log N)$ operation in the worst case.
During unwinding, any nodes marked as dirty are reconstructed to incorporate pending modifications.
Importantly, \verb+value+ is a pure operation: it does not invalidate or alter the iterator state.

\myparagraph{Example}
Extracting the value from the zipper \textcolor{red!50!black}{\circled{Y}} from \autoref{fig:zipper} will propagate the $\dirtyflag$ updates and reconstruct a finger tree over the full sequence $\langle 1, 2, ..., 29 \rangle$ including the previously missing element \verb+9+.

\subsubsection{Zippers as a Unifying Iterator Abstraction}

As with persistent containers, our persistent iterator design is uniform across all common container types.
Since every container in the library ({\em vectors}, {\em sets}, {\em maps}, and {\em strings}) is ultimately a specialization of the same finger tree abstraction, all iterators share a single implementation.

\section{Optimization}

Although persistent data structures offer strong semantic and algorithmic advantages, they can easily become allocation-bound.
This is because each update typically requires creating and releasing $O(\log N)$  nodes, and this cost may dominate execution time for some workloads.
Moreover, tree-based data structures are inherently non-contiguous in memory, which can harm locality.
These issues are the primary reason behind the ``persistence tax''.
This section describes the low-level optimizations used by \tool to help minimize this tax.

\subsection{A Very Fast Custom Memory Allocator (CMA)}

Our first optimization tackles the cost of allocation by designing a specialized allocator and deallocator that execute in just a few machine instructions.
Our design is based on two key insights.

\myparagraph{Uniform Node Size}
The first insight is that all finger-tree and zipper nodes in \tool have roughly uniform size and fit within a single 64-byte cache line.
This allows us to implement a {\em fixed-size} free-list allocator, where: {\em allocation} pops an entry from the head of the free-list, and {\em deallocation} pushes the node back.
Because no variable-size blocks need to be handled, the allocator is trivially simple and efficient.

\myparagraph{Zero-initialized Memory as a Linked List}
The second insight is that zeroed memory can itself encode a valid linked list. Instead of storing a raw pointer to the {\em next} free node, each node stores the {\em offset} between the next free node (in the list) and the next node in memory.
Thus, if the next free node (in the list) happens to be the next free node (in memory), the offset will be \texttt{0x0}.

This means that a zeroed memory region is already a valid free-list over its length, without any explicit initialization pass.
The {\em heap} for the CMA is created with a single call to \verb+mmap+ with the \verb+NO_RESERVE+ flag---ensuring that the heap is zero'ed and {\em lazily committed} and grows by demand as the address space expands.
Furthermore, heap allocation never \emph{fails} in the traditional sense, i.e., by returning \verb+NULL+.
Rather, heap allocation behaves conceptually like stack allocation over a bounded region: physical memory usage is based on demand.
The resulting allocator is {\em extremely compact} for single-threaded code, and (de)allocation is reduced to a handful of instructions (see \autoref{fig:cma}).
Our implementation supports different allocators, including a multi-threaded variant of the CMA, as well as wrappers for the standard allocators \verb|new| and \verb|delete|.

\begin{figure}
\lstdefinelanguage{asm}{
  morekeywords={mov,lea,sub,add,push,pop,ret,call,jmp,cmp,test,je,jne,jz,jnz},
  sensitive=true,
  morecomment=[l]{\#},
  morestring=[b]",
}

\lstdefinestyle{asmstyle}{
  language=asm,
  basicstyle=\ttfamily\scriptsize,
  keywordstyle=\color{blue},
  morekeywords={string, list, char32_t},
  keywordstyle=[2]\color{teal},
  commentstyle=\color{gray}\itshape,
  stringstyle=\color{red},
  showstringspaces=false,
  numbers=none,
  escapeinside={(*@}{@*)},
  tabsize=2,
  xleftmargin=1em,
  frame=single,
  framesep=2pt,
  backgroundcolor=\color{gray!8},
}

\begin{minipage}{0.45\textwidth}
\begin{lstlisting}[style=asmstyle]
# allocation (result in %rax)

mov HEAP, %rbx
mov (%rbx), %rax
lea 64(%rbx, %rax), %rax
mov %rax, HEAP
\end{lstlisting}
\end{minipage}
$\quad$
\begin{minipage}{0.45\textwidth}
\begin{lstlisting}[style=asmstyle]
# deallocation (node pointer in %rax)
mov HEAP, %rbx
sub %rax, %rbx
sub $64, %rbx
mov %rbx, (%rax)
mov %rax, HEAP
\end{lstlisting}
\end{minipage}
\caption{Highly optimized (de)allocation routines (\texttt{x86\_64}, single-threaded), with 4 inlined instructions for allocation, and 5 inlined instructions for deallocation.
\label{fig:cma}}
\end{figure}

\subsection{Other Optimizations}

While node allocation is extremely cheap, it is also not free.
\tool therefore performs {\em destructive updates} whenever it is safe to do so---specifically, when all nodes along an update path have a reference count of exactly one.
In such cases, the affected nodes are updated {\em in-place} rather than copied.
\tool applies this to many common operations, including iterator increment \verb_++it_.
In practice, a large fraction of nodes meet the ``unique reference'' condition, so a destructive update can eliminate the majority of allocations in typical workloads.
Although not novel (also implemented by Immer~\cite{puente17immer} and other systems), this optimization helps minimize algorithmic constant factors.

Finally, \tool packs multiple elements into each leaf node, as much as possible to fill an entire cache line.
This reduces tree depth, improves spatial locality, and minimizes traversal overhead.
Fat leaves are especially beneficial for sequential scans, where they exploit hardware prefetching and contiguous memory access patterns.

\section{Evaluation}
To evaluate the effectiveness of \tool persistent containers/iterators, we consider the following:

\begin{itemize}[leftmargin=*]
\item[] \textbf{(\hyperref[sec:rq1]{Algorithmic Complexity})} What is the algorithmic complexity of \tool container operations compared to the baseline?
\item[] \textbf{(\hyperref[sec:rq2]{Constant Factors})} What are the constant factors of \tool container operations?
\item[] \textbf{(\hyperref[sec:rq3]{Memory Overheads})} What is the memory overhead of \tool containers?
\end{itemize}
For the constant factor and memory overheads, we focus on operation-level micro-benchmarks to isolate the cost of persistence and functional updates.

\subsection{Experimental Setup}

As \tool is intended to act as a persistent container alternative for general container and iterator usages, we compare it against several popular libraries in this space as baselines. Specifically, we compared against the following libraries:

\begin{itemize}[leftmargin=*]
\item \emph{Standard Template Library} (\textsc{STL}) (\verb_libc++_ 21.0.0): Containers and iterators have been present since the first iteration of the Standard Template Library back in 1995~\cite{stepanov1995stl}.
Due to its inclusion in the \verb_C++_ standard library, commonly shipped together with popular \verb_C++_ compilers, the STL has become the default@b for container and iterator usage in \verb_C++_ programming.

\item \textsc{Immer} (0.8.1): Immer~\cite{puente17immer} is an alternative persistent container library realizing {\em Relaxed Radix Balanced} (RRB) trees for sequence-like containers {\em Compressed Hash-Array Mapped Prefix-tree} (CHAMP) structures~\cite{steindorfer15champ} for maps/sets.
Unlike \tool, Immer only supports {\em const} rather than {\em persistent iterators}, meaning that it is not possible to update containers via iteration.

\item \textsc{Abseil} (20250814.1): Abseil~\cite{Abseil} is a popular open-source collection of \verb_C++_ library code designed to augment the \verb_C++_ standard library, containing a significant number of useful functionalities used internally at Google that are not present in the STL.
Alternative containers also form a significant component of Abseil, including Swiss table versions of unordered containers, such as \texttt{absl::flat\_hash\_map}, and B-Tree versions of ordered containers, such as \texttt{absl::btree\_map}.
Similar to \tool, Abseil containers are designed to be replacements for their STL counterparts, following the standard container API if possible.
Unlike \tool and Immer, Abseil containers are not persistent.

\item \textsc{Folly} (2025.11.03.00): Folly~\cite{Folly} (Facebook Open Source Library) is another popular open-source \verb_C++_ library with components designed with practicality and efficiency in mind.
Similar to Abseil, Folly is essentially a collection of useful internal components extensively deployed inside Meta, and only strives to complement the functionalities of the STL instead of replacing it.
Containers also form a significant component of Folly, including several items that the standard library neglects to provide, such as high-performance atomic data structures, a drop-in replacement version of \texttt{std::string} and \texttt{std::vector}, intrusive linked lists, and more.
Unlike \tool and Immer, Folly containers are not persistent.
\end{itemize}
Finally, as the high-performance thread-local CMA is used in the default implementation of \tool, an alternative version marked as ``\tool (unopt)" is also evaluated with the standard allocator instead to reduce non-algorithmic impact on performance evaluations.

\subsection{Algorithmic Complexity} \label{sec:rq1}

To evaluate the theoretical efficiency of \tool containers, we performed a complexity analysis of all core operations and compared them with their counterparts in the \verb_C++_ {\em Standard Template Library} (STL).
The analysis is based on the known asymptotic properties of the persistent data structures underlying \tool, namely, finger trees and zippers.
For each operation, we consider either the {\em amortized} or {\em worst-case} cost, depending on the semantics of the corresponding STL operation.
The results are summarized in \autoref{tab:complexity}.

\subsubsection{Results}

\tool provides persistent counterparts to nearly all common STL container and iterator operations, while preserving similar syntax and equivalent time/space complexities in most cases.
This equivalence is by design: finger trees achieve $O(1)$ amortized append and access to the ends of a sequence, and zippers support traversal in constant time increments, yielding asymptotic behavior comparable to standard mutable containers.

That said, there are some notable differences arising from the fundamental distinction between array-based and tree-based representations.
For example, random access \verb+v[idx]+ and arbitrary iterator jumps (\verb_it += n_) require $O(\log N)$ time in \tool, whereas contiguous STL vectors perform the same operations in $O(1)$.
This is a direct consequence of persistence: maintaining previous versions efficiently requires node-based structural sharing rather than contiguous arrays.
Such logarithmic penalties are typical of all general-purpose persistent sequences.

Conversely, several operations benefit from the persistent design.
Element insertion and erasure through iterators (\verb+it.insert(x)+, \verb+it.erase()+) are amortized $O(1)$ operations under \tool, compared to $O(N)$ or their STL vector equivalents.
String concatenation and splitting are likewise logarithmic rather than linear, as finger trees support efficient join and split operations.
Sequence containers in \tool naturally support both \verb+push_front+ and \verb+push_back+ in amortized $O(1)$ time, similar to \verb+std::deque+ but without additional container specialization.
The results from \autoref{tab:complexity} also show that the complexity for {\em any} standard operation is no more than $O(\log N)$.
This reduces the risk of ``complexity hazards'' when implementing natural versions of standard algorithms, such as those that exist in the STL (e.g., see \autoref{fig:examples}~(b) over \verb|std::string|).

\begin{table*}[htbp]
\centering
\scriptsize
\setlength{\tabcolsep}{3pt}
\renewcommand{\arraystretch}{1.2}
\begin{tabular}{llllll}
\toprule
\textbf{STL Operation} & \textbf{STL Time} & \textbf{STL Space} & \textbf{\tool Operation} & \textbf{\tool Time} & \textbf{\tool Space} \\
\midrule
\multicolumn{6}{c}{\textbf{Vector}} \\
\midrule
\texttt{v.size()} & $\mathcal{O}(1)$ & $\mathcal{O}(1)$ & \texttt{v.size()} & \cellcolor{yellow!20}$\mathcal{O}(1)$ & \cellcolor{yellow!20}$\mathcal{O}(1)$ \\
\texttt{v[idx]} & $\mathcal{O}(1)$ & $\mathcal{O}(1)$ & \texttt{v[idx]} & \cellcolor{red!20}$\mathcal{O}(\log N)$ & \cellcolor{yellow!20}$\mathcal{O}(1)$ \\
\texttt{v.back()} & $\mathcal{O}(1)$ & $\mathcal{O}(1)$ & \texttt{v.back()} & \cellcolor{yellow!20}$\mathcal{O}(1)$ & \cellcolor{yellow!20}$\mathcal{O}(1)$ \\
\texttt{v.front()} & $\mathcal{O}(1)$ & $\mathcal{O}(1)$ & \texttt{v.front()} & \cellcolor{yellow!20}$\mathcal{O}(1)$ & \cellcolor{yellow!20}$\mathcal{O}(1)$ \\
\texttt{v.push\_back(x)} & $\mathcal{O}(1)$ & $\mathcal{O}(1)$ & \texttt{v.push\_back(x)} & \cellcolor{yellow!20}$\mathcal{O}(1)$ & \cellcolor{yellow!20}$\mathcal{O}(1)$ \\
\hspace{1em}{\em n/a} & - & - & \texttt{v.push\_front(x)} & \cellcolor{green!0}$\mathcal{O}(1)$ & \cellcolor{green!0}$\mathcal{O}(1)$ \\
\texttt{v.pop\_back()} & $\mathcal{O}(1)$ & $\mathcal{O}(1)$ & \texttt{v.pop\_back()} & \cellcolor{yellow!20}$\mathcal{O}(1)$ & \cellcolor{yellow!20}$\mathcal{O}(1)$ \\
\hspace{1em}{\em n/a} & - & - & \texttt{v.pop\_front()} & \cellcolor{green!0}$\mathcal{O}(1)$ & \cellcolor{green!0}$\mathcal{O}(1)$ \\
\texttt{it = v.begin()} & $\mathcal{O}(1)$ & $\mathcal{O}(1)$ & \texttt{it = v.begin()} & \cellcolor{yellow!20}$\mathcal{O}(1)$ & \cellcolor{yellow!20}$\mathcal{O}(1)$ \\
\texttt{it = v.insert(it,x)} & $\mathcal{O}(N)$ & $\mathcal{O}(1)$ & \texttt{it.insert(x)} & \cellcolor{green!40}$\mathcal{O}(1)$ & \cellcolor{yellow!20}$\mathcal{O}(1)$ \\
\texttt{it = v.erase(it)} & $\mathcal{O}(N)$ & $\mathcal{O}(1)$ & \texttt{it.erase()} & \cellcolor{green!40}$\mathcal{O}(1)$ & \cellcolor{yellow!20}$\mathcal{O}(1)$ \\
\texttt{*it = x} & $\mathcal{O}(1)$ & $\mathcal{O}(1)$ & \texttt{it.assign(x)} & \cellcolor{yellow!20}$\mathcal{O}(1)$ & \cellcolor{yellow!20}$\mathcal{O}(1)$ \\
\texttt{++it}/\texttt{{-}{-}it}/\texttt{*it} & $\mathcal{O}(1)$ & $\mathcal{O}(1)$ & \texttt{++it}/\texttt{{-}{-}it}/\texttt{*it} & \cellcolor{yellow!20}$\mathcal{O}(1)$ & \cellcolor{yellow!20}$\mathcal{O}(1)$ \\
\texttt{it += n} & $\mathcal{O}(1)$ & $\mathcal{O}(1)$ & \texttt{it += n} & \cellcolor{red!20}$\mathcal{O}(\log N)$ & \cellcolor{yellow!20}$\mathcal{O}(1)$ \\
\midrule
\multicolumn{6}{c}{\textbf{Set} and \textbf{Map}} \\
\midrule
\texttt{m.size()} & $\mathcal{O}(1)$ & $\mathcal{O}(1)$ & \texttt{m.size()} & \cellcolor{yellow!20}$\mathcal{O}(1)$ & \cellcolor{yellow!20}$\mathcal{O}(1)$ \\
\texttt{m[k]} & $\mathcal{O}(\log N)$ & $\mathcal{O}(1)$ & \texttt{m[k]} & \cellcolor{yellow!20}$\mathcal{O}(\log N)$ & \cellcolor{yellow!20}$\mathcal{O}(1)$ \\
\hspace{1em}{\em n/a} & - & - & \texttt{m[idx]} & \cellcolor{green!0}$\mathcal{O}(\log N)$ & \cellcolor{green!0}$\mathcal{O}(1)$ \\
\texttt{m.contains(k)} & $\mathcal{O}(\log N)$ & $\mathcal{O}(1)$ & \texttt{m.contains(k)} & \cellcolor{yellow!20}$\mathcal{O}(\log N)$ & \cellcolor{yellow!20}$\mathcal{O}(1)$ \\
\texttt{m.insert(x)} & $\mathcal{O}(\log N)$ & $\mathcal{O}(1)$ & \texttt{m.insert(x)} & \cellcolor{yellow!20}$\mathcal{O}(\log N)$ & \cellcolor{yellow!20}$\mathcal{O}(1)$ \\
\texttt{m.erase(x)} & $\mathcal{O}(\log N)$ & $\mathcal{O}(1)$ & \texttt{m.erase(x)} & \cellcolor{yellow!20}$\mathcal{O}(\log N)$ & \cellcolor{yellow!20}$\mathcal{O}(1)$ \\
\texttt{it = m.begin()} & $\mathcal{O}(1)$ & $\mathcal{O}(1)$ & \texttt{it = m.begin()} & \cellcolor{yellow!20}$\mathcal{O}(1)$ & \cellcolor{yellow!20}$\mathcal{O}(1)$ \\
\texttt{it = m.find(x)} & $\mathcal{O}(\log N)$ & $\mathcal{O}(1)$ & \texttt{it = m.find(x)} & \cellcolor{yellow!20}$\mathcal{O}(\log N)$ & \cellcolor{red!20}$\mathcal{O}(\log N)$ \\
\texttt{it = m.erase(it)} & $\mathcal{O}(1)$ & $\mathcal{O}(1)$ & \texttt{it.erase()} & \cellcolor{yellow!20}$\mathcal{O}(1)$ & \cellcolor{yellow!20}$\mathcal{O}(1)$ \\
\texttt{++it}/\texttt{{-}{-}it}/\texttt{*it} & $\mathcal{O}(1)$ & $\mathcal{O}(1)$ & \texttt{++it}/\texttt{{-}{-}it}/\texttt{*it} & \cellcolor{yellow!20}$\mathcal{O}(1)$ & \cellcolor{yellow!20}$\mathcal{O}(1)$ \\
\hspace{1em}{\em n/a} & - & - & \texttt{it += n} & \cellcolor{green!0}$\mathcal{O}(\log N)$ & \cellcolor{green!0}$\mathcal{O}(1)$ \\
\midrule
\multicolumn{6}{c}{\textbf{String}} \\
\ignore{
\multicolumn{6}{c}{\textbf{Map}} \\
\midrule
\texttt{m.size()} & $\mathcal{O}(1)$ & $\mathcal{O}(1)$ & \texttt{xs.size()} & \cellcolor{yellow!20}$\mathcal{O}(1)$ & \cellcolor{yellow!20}$\mathcal{O}(1)$ \\
\texttt{m.clear()} & $\mathcal{O}(N)$ & $\mathcal{O}(1)$ & \texttt{xs.clear()} & \cellcolor{green!20}$\mathcal{O}(1)$ & \cellcolor{green!20}$\mathcal{O}(1)$ \\
\texttt{m.find(k)} & $\mathcal{O}(\log N)$ & $\mathcal{O}(1)$ & \texttt{xs.find(k)} & \cellcolor{yellow!20}$\mathcal{O}(\log N)$ & \cellcolor{yellow!20}$\mathcal{O}(1)$ \\
\texttt{m.contains(k)} & $\mathcal{O}(\log N)$ & $\mathcal{O}(1)$ & \texttt{xs.contains(k)} & \cellcolor{yellow!20}$\mathcal{O}(\log N)$ & \cellcolor{yellow!20}$\mathcal{O}(1)$ \\
\texttt{m.insert(k,v)} & $\mathcal{O}(\log N)$ & $\mathcal{O}(1)$ & \texttt{xs.insert(k,v)} & \cellcolor{yellow!20}$\mathcal{O}(\log N)$ & \cellcolor{yellow!20}$\mathcal{O}(\log N)$ \\
\texttt{m.erase(k)} & $\mathcal{O}(\log N)$ & $\mathcal{O}(1)$ & \texttt{xs.erase(k)} & \cellcolor{yellow!20}$\mathcal{O}(\log N)$ & \cellcolor{yellow!20}$\mathcal{O}(\log N)$ \\
\texttt{m[k]} & $\mathcal{O}(\log N)$ & $\mathcal{O}(1)$ & \texttt{xs[k]} & \cellcolor{yellow!20}$\mathcal{O}(\log N)$ & \cellcolor{yellow!20}$\mathcal{O}(\log N)$ \\
\texttt{m.at(k)} & $\mathcal{O}(\log N)$ & $\mathcal{O}(1)$ & \texttt{xs.at(k)} & \cellcolor{yellow!20}$\mathcal{O}(\log N)$ & \cellcolor{yellow!20}$\mathcal{O}(\log N)$ \\
\midrule
}
\midrule
\texttt{s.size()} & $\mathcal{O}(1)$ & $\mathcal{O}(1)$ & \texttt{s.size()} & \cellcolor{yellow!20}$\mathcal{O}(1)$ & \cellcolor{yellow!20}$\mathcal{O}(1)$ \\
\texttt{s[idx]} & $\mathcal{O}(1)$ & $\mathcal{O}(1)$ & \texttt{s[idx]} & \cellcolor{red!20}$\mathcal{O}(\log N)$ & \cellcolor{yellow!20}$\mathcal{O}(1)$ \\
\texttt{s += c} & $\mathcal{O}(1)$ & $\mathcal{O}(1)$ & \texttt{s += c} & \cellcolor{yellow!20}$\mathcal{O}(1)$ & \cellcolor{yellow!20}$\mathcal{O}(1)$ \\
\hspace{1em}{\em n/a} & - & - & \texttt{s.push\_front(c)} & \cellcolor{green!0}$\mathcal{O}(1)$ & \cellcolor{green!0}$\mathcal{O}(1)$ \\
\texttt{s += t} & $\mathcal{O}(N+M)$ & $\mathcal{O}(N+M)$ & \texttt{s += t} & \cellcolor{green!40}$\mathcal{O}(\min(\log N, \log M))$ & \cellcolor{green!40}$\mathcal{O}(\min(\log N, \log M))$ \\
\texttt{s.substr(idx)} & $\mathcal{O}(N)$ & $\mathcal{O}(N)$ & \texttt{s.substr(idx)} & \cellcolor{green!40}$\mathcal{O}(\log N)$ & \cellcolor{green!40}$\mathcal{O}(\log N)$ \\
\texttt{it = s.begin()} & $\mathcal{O}(1)$ & $\mathcal{O}(1)$ & \texttt{it = s.begin()} & \cellcolor{yellow!20}$\mathcal{O}(1)$ & \cellcolor{yellow!20}$\mathcal{O}(1)$ \\
\texttt{it = s.insert(it,c)} & $\mathcal{O}(N)$ & $\mathcal{O}(1)$ & \texttt{it.insert(c)} & \cellcolor{green!40}$\mathcal{O}(1)$ & \cellcolor{yellow!20}$\mathcal{O}(1)$ \\
\texttt{it = s.erase(it)} & $\mathcal{O}(N)$ & $\mathcal{O}(1)$ & \texttt{it.erase()} & \cellcolor{green!40}$\mathcal{O}(1)$ & \cellcolor{yellow!20}$\mathcal{O}(1)$ \\
\texttt{*it = c} & $\mathcal{O}(1)$ & $\mathcal{O}(1)$ & \texttt{it.assign(c)} & \cellcolor{yellow!20}$\mathcal{O}(1)$ & \cellcolor{yellow!20}$\mathcal{O}(1)$ \\
\texttt{++it}/\texttt{{-}{-}it}/\texttt{*it} & $\mathcal{O}(1)$ & $\mathcal{O}(1)$ & \texttt{++it}/\texttt{{-}{-}it}/\texttt{*it} & \cellcolor{yellow!20}$\mathcal{O}(1)$ & \cellcolor{yellow!20}$\mathcal{O}(1)$ \\
\texttt{it += n} & $\mathcal{O}(1)$ & $\mathcal{O}(1)$ & \texttt{it += n} & \cellcolor{red!20}$\mathcal{O}(\log N)$ & \cellcolor{yellow!20}$\mathcal{O}(1)$ \\
\bottomrule
\end{tabular}
\caption{Comparison of time and space complexities between the {\em Standard Template Library} (STL) and \tool containers.
Key:
\texttt{v}${=}${\em vector}, \texttt{m}${=}${\em map}/{\em set}, \texttt{s}/\texttt{t}${=}${\em strings}, \texttt{idx}/\texttt{n}${=}${\em integers}, \texttt{x}${=}${\em value}, \texttt{k}${=}${\em key}, \texttt{c}${=}${\em character}, \colorbox{green!40}{green}${=}$better, \colorbox{yellow!20}{yellow}${=}$same, \colorbox{red!20}{red}${=}$worse.\label{tab:complexity}}
\end{table*}

\result{\tool achieves time and space complexities comparable to the STL across nearly all operations while adding full value semantics without complexity hazards.
In cases where asymptotic costs differ, the trade-offs stem from persistence requirements rather than inefficiency in design.
\vspace{0.5em}}

\subsection{Constant Factors} \label{sec:rq2}

While asymptotic complexity provides a useful theoretical baseline, the practical performance of a container implementation also depends on the associated constant factors.
Persistent data structures are inherently disadvantaged in this regard: structural sharing and indirection reduce spatial locality, and operations such as updates and traversal often involve pointer chasing across disjoint memory regions.  Consequently, we do not expect persistent containers to outperform their mutable counterparts.
Nevertheless, the following section focuses on evaluating constant-factor effects through targeted micro-benchmarks to quantify the impact of persistence.

\subsubsection{Containers}\label{sec:containers}
All experiments were carried out on an macOS \verb+arm64+ machine operating on M2 Pro with 32 GB of memory, compiled with Clang 21.1 released in August 2025.
The benchmarks in these experiments use the popular testing framework Catch2~\cite{Catch2} (3.11.0), which provides comprehensive unit-testing for \tool itself.
To evaluate the constant factors associated with \tool's persistent containers, we use the four micro-benchmarks adapted from ~\cite{puente17immer}.

\begin{itemize}[leftmargin=*]
\item \textsc{Access}: The sum of all values in a $n$ element container is computed  by sequential indexes;
\item \textsc{Append}: An $n$ element container is produced by sequentially appending $n$ elements;
\item \textsc{Update}: Every element in an $n$ element container is updated using sequential indexes;
\item \textsc{Concat}: An $n$ element container is produced by concatenating 10 equally sized containers.
\end{itemize}
We evaluate {\em vectors} as a representative of sequence-based containers ({\em vectors}, {\em strings}, etc.), and {\em sets} as a representative of node-based containers ({\em multisets}, {\em maps}, etc.).
Note that \textsc{Update} and \textsc{Concat} are for sequence-based containers only.

\begin{figure}
\centering
\begin{tikzpicture}[scale=0.6]
\begin{axis}[
    width=7.4cm,
    height=5cm,
    xlabel={Size},
    ylabel={Time (ns)},
    title={{\sc Access} ({\tt vector})},
    title style={yshift=-1.5ex},
    grid=none,
    grid style={white!30!black},
    legend style={
        at={(0.03,0.97)},
        anchor=north west,
        draw=none,
        fill=none,
        text=black,
        font=\tiny,
        cells={anchor=west},
        draw=black,
        fill=white,
    },
    ticklabel style={text=black},
    label style={font=\small, yshift=0.5ex, text=black},
    title style={text=black},
    axis background/.style={fill=white},
    every axis plot/.append style={thick},
    xmode=log,
    ymode=log,
]

\addplot[
    color=darkgreen,
] table [x=Size, y=F::vector, col sep=comma] {data/access-vector.csv};

\addlegendentry{\tool}

\addplot[
    color=darkgreen, dashed,
] table [x=Size, y=F::vector, col sep=comma] {data/access-vector-slow.csv};

\addlegendentry{\tool (unopt)}

\addplot[
    color=red,
] table [x=Size, y=std::vector, col sep=comma] {data/access-vector.csv};
\addlegendentry{STL}

\addplot[
    color=cyan,
] table [x=Size, y=absl::InlinedVector, col sep=comma] {data/access-vector.csv};
\addlegendentry{\sc Abseil}

\addplot[
    color=blue,
] table [x=Size, y=folly::fbvector, col sep=comma] {data/access-vector.csv};
\addlegendentry{\sc Folly}

\addplot[
    color=magenta, densely dotted,
] table [x=Size, y=immer::flex_vector, col sep=comma] {data/access-vector.csv};
\addlegendentry{\sc Immer}

\end{axis}
\end{tikzpicture}
\quad
\begin{tikzpicture}[scale=0.6]
\begin{axis}[
    width=7.4cm,
    height=5cm,
    xlabel={Size},
    ylabel={Time (ns)},
    title={{\sc Append} ({\tt vector})},
    title style={yshift=-1.5ex},
    grid=none,
    grid style={white!30!black},
    legend style={
        at={(0.03,0.97)},
        anchor=north west,
        draw=none,
        fill=none,
        text=black,
        font=\footnotesize,
        cells={anchor=west},
        draw=black,
        fill=white,
    },
    ticklabel style={text=black},
    label style={font=\small, yshift=0.5ex, text=black},
    title style={text=black},
    axis background/.style={fill=white},
    every axis plot/.append style={thick},
    xmode=log,
    ymode=log,
]

\addplot[
    color=darkgreen,
] table [x=Size, y=F::vector, col sep=comma] {data/append-vector.csv};

\addplot[
    color=darkgreen, dashed,
] table [x=Size, y=F::vector, col sep=comma] {data/append-vector-slow.csv};

\addplot[
    color=red,
] table [x=Size, y=std::vector, col sep=comma] {data/append-vector.csv};

\addplot[
    color=cyan,
] table [x=Size, y=absl::InlinedVector, col sep=comma] {data/append-vector.csv};

\addplot[
    color=blue,
] table [x=Size, y=folly::fbvector, col sep=comma] {data/append-vector.csv};

\addplot[
    color=magenta, densely dotted,
] table [x=Size, y=immer::flex_vector, col sep=comma] {data/append-vector.csv};

\end{axis}
\end{tikzpicture}
\quad
\begin{tikzpicture}[scale=0.6]
\begin{axis}[
    width=7.4cm,
    height=5cm,
    xlabel={Size},
    ylabel={Time (ns)},
    title={{\sc Update} ({\tt vector})},
    title style={yshift=-1.5ex},
    grid=none,
    grid style={white!30!black},
    legend style={
        at={(0.03,0.97)},
        anchor=north west,
        draw=none,
        fill=none,
        text=black,
        font=\footnotesize,
        cells={anchor=west},
        draw=black,
        fill=white,
    },
    ticklabel style={text=black},
    label style={font=\small, yshift=0.5ex, text=black},
    title style={text=black},
    axis background/.style={fill=white},
    every axis plot/.append style={thick},
    xmode=log,
    ymode=log,
]

\addplot[
    color=darkgreen,
] table [x=Size, y=F::vector, col sep=comma] {data/update-vector.csv};

\addplot[
    color=darkgreen, dashed,
] table [x=Size, y=F::vector, col sep=comma] {data/update-vector-slow.csv};

\addplot[
    color=red,
] table [x=Size, y=std::vector, col sep=comma] {data/update-vector.csv};

\addplot[
    color=cyan,
] table [x=Size, y=absl::InlinedVector, col sep=comma] {data/update-vector.csv};

\addplot[
    color=blue,
] table [x=Size, y=folly::fbvector, col sep=comma] {data/update-vector.csv};

\addplot[
    color=magenta, densely dotted,
] table [x=Size, y=immer::flex_vector, col sep=comma] {data/update-vector.csv};

\end{axis}
\end{tikzpicture}

\begin{tikzpicture}[scale=0.6]
\begin{axis}[
    width=7.4cm,
    height=5cm,
    xlabel={Size},
    ylabel={Time (ns)},
    title={{\sc Concat} ({\tt vector})},
    title style={yshift=-1.5ex},
    grid=none,
    grid style={white!30!black},
    legend style={
        at={(0.03,0.97)},
        anchor=north west,
        draw=none,
        fill=none,
        text=black,
        font=\footnotesize,
        cells={anchor=west},
        draw=black,
        fill=white,
    },
    ticklabel style={text=black},
    label style={font=\small, yshift=0.5ex, text=black},
    title style={text=black},
    axis background/.style={fill=white},
    every axis plot/.append style={thick},
    xmode=log,
    ymode=log,
]

\addplot[
    color=darkgreen,
] table [x=Size, y=F::vector, col sep=comma] {data/concat-vector.csv};

\addplot[
    color=darkgreen, dashed,
] table [x=Size, y=F::vector, col sep=comma] {data/concat-vector-slow.csv};

\addplot[
    color=red,
] table [x=Size, y=std::vector, col sep=comma] {data/concat-vector.csv};

\addplot[
    color=cyan,
] table [x=Size, y=absl::InlinedVector, col sep=comma] {data/concat-vector.csv};

\addplot[
    color=blue,
] table [x=Size, y=folly::fbvector, col sep=comma] {data/concat-vector.csv};

\addplot[
    color=magenta, densely dotted,
] table [x=Size, y=immer::flex_vector, col sep=comma] {data/concat-vector.csv};

\end{axis}
\end{tikzpicture}
\quad
\begin{tikzpicture}[scale=0.6]
\begin{axis}[
    width=7.4cm,
    height=5cm,
    xlabel={Size},
    ylabel={Time (ns)},
    title={{\sc Access} ({\tt set})},
    title style={yshift=-1.5ex},
    grid=none,
    grid style={white!30!black},
    legend style={
        at={(0.03,0.97)},
        anchor=north west,
        draw=none,
        fill=none,
        text=black,
        font=\footnotesize,
        cells={anchor=west},
        draw=black,
        fill=white,
    },
    ticklabel style={text=black},
    label style={font=\small, yshift=0.5ex, text=black},
    title style={text=black},
    axis background/.style={fill=white},
    every axis plot/.append style={thick},
    xmode=log,
    ymode=log,
]

\addplot[
    color=darkgreen,
] table [x=Size, y=F::set, col sep=comma] {data/access-set.csv};

\addplot[
    color=darkgreen, dashed,
] table [x=Size, y=F::set, col sep=comma] {data/access-set-slow.csv};

\addplot[
    color=red,
] table [x=Size, y=std::set, col sep=comma] {data/access-set.csv};

\addplot[
    color=cyan,
] table [x=Size, y=absl::btree_set, col sep=comma] {data/access-set.csv};

\addplot[
    color=blue,
] table [x=Size, y=folly::F14FastSet, col sep=comma] {data/access-set.csv};

\addplot[
    color=magenta, densely dotted,
] table [x=Size, y=immer::set, col sep=comma] {data/access-set.csv};

\end{axis}
\end{tikzpicture}
\quad
\begin{tikzpicture}[scale=0.6]
\begin{axis}[
    width=7.4cm,
    height=5cm,
    xlabel={Size},
    ylabel={Time (ns)},
    title={{\sc Append} ({\tt set})},
    title style={yshift=-1.5ex},
    grid=none,
    grid style={white!30!black},
    legend style={
        at={(0.03,0.97)},
        anchor=north west,
        draw=none,
        fill=none,
        text=black,
        font=\footnotesize,
        cells={anchor=west},
        draw=black,
        fill=white,
    },
    ticklabel style={text=black},
    label style={font=\small, yshift=0.5ex, text=black},
    title style={text=black},
    axis background/.style={fill=white},
    every axis plot/.append style={thick},
    xmode=log,
    ymode=log,
]

\addplot[
    color=darkgreen,
] table [x=Size, y=F::set, col sep=comma] {data/append-set.csv};

\addplot[
    color=darkgreen, dashed,
] table [x=Size, y=F::set, col sep=comma] {data/append-set-slow.csv};

\addplot[
    color=red,
] table [x=Size, y=std::set, col sep=comma] {data/append-set.csv};

\addplot[
    color=cyan,
] table [x=Size, y=absl::btree_set, col sep=comma] {data/append-set.csv};

\addplot[
    color=blue,
] table [x=Size, y=folly::F14FastSet, col sep=comma] {data/append-set.csv};

\addplot[
    color=magenta, densely dotted,
] table [x=Size, y=immer::set, col sep=comma] {data/append-set.csv};

\end{axis}
\end{tikzpicture}

\caption{Figure showing container micro-benchmark results for {\em vectors} and {\em sets} in \tool and other libraries under different container sizes.  Scales are logarithmic.
Compared to the STL baseline, \tool is $56.9{\times}$/$3.4{\times}$/$262.0{\times}$ slower for the \textsc{Access}/\textsc{Append}/\textsc{Update} (\texttt{vector}) benchmarks,
$3.3{\times}$ slower for the \textsc{Access} benchmark, $7.6{\times}$ faster for the \textsc{Concat} (\texttt{vector}) benchmark, and $1.8{\times}$ faster for the \textsc{Append} (\texttt{set}) benchmark.
\label{fig:vector-micro}}
\end{figure}
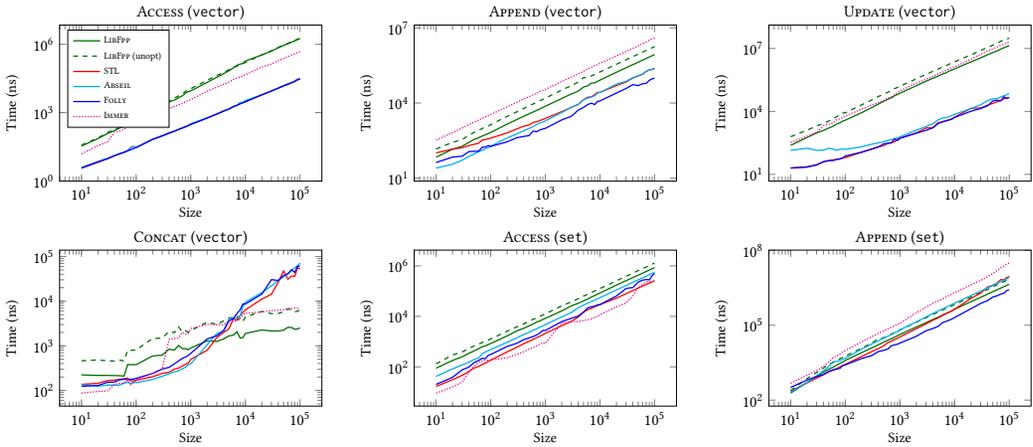

\subsubsection*{Results}

The micro-benchmark results are summarized in \autoref{fig:vector-micro}.
As expected, the performance characteristics of \tool are broadly similar to those of Immer, since both employ persistent tree-based representations with structural sharing.
For sequence-based containers, however, the mutable counterparts retain a substantial advantage due to their {\em array-based} implementations.
For example, index-based access \verb|v[idx]| on a contiguous array is generally compiled into just a few machine instructions, and also benefits from excellent cache locality.
This difference in representation results in a $3.4\times$ slowdown for the \textsc{Append} (\texttt{vector}) benchmark, where both \tool and STL implement an amortized $O(1)$ operation.
For \textsc{Access} and \textsc{Update} (\texttt{vector}), both persistent implementations must traverse a tree structure to locate the element ($O(\log n)$ versus $O(1)$), resulting in higher observed overheads in the order of 10${\times}$-100${\times}$ or more.
For \textsc{Concat} (\texttt{vector}) the opposite occurs, where the tree-based representation offers an algorithmic advantage over array-based containers, resulting in a speedup.
These cost differences are fundamental to persistence and are not easily eliminated without sacrificing immutability.

Within the persistent domain, the design trade-offs of \tool and \textit{Immer} become more apparent.
Immer employs {\em Relaxed Radix Balanced} (RRB)~\cite{stucki15rrb} trees for sequence containers and {\em Compressed Hash-Array Mapped Prefix-tree} (CHAMP) structures~\cite{steindorfer15champ} for maps/sets, each with a large node fan-out (typically up to 32).
\tool, by contrast, uses a uniform finger-tree foundation with a fixed node size of 64 bytes.
This distinction is reflected in the results: \tool shows higher constant-factor overheads for indexed benchmarks such as \textsc{Access}, but achieves lower costs for modification-heavy workloads such as \textsc{Append}, \textsc{Update}, and \textsc{Concat}, regardless of allocator choice.
The performance gap is primarily attributable to differences in high fan-out RRBs/CHAMPs versus finger trees, with the former allowing for shallower lookups at the cost of more expensive updates.

\subsubsection{Iterators}
To test persistent iterators, we consider the following modified micro-benchmarks:
\begin{itemize}[leftmargin=*]
\item \textsc{Access}$^*$: The sum of all values in a $n$ element container is computed by sequential iteration;
\item \textsc{Append}$^*$: An $n$ element container is produced by appending $n$ elements via an iterator;
\item \textsc{Update}$^*$: Every element in an $n$ element container is updated via an iterator;
\item \textsc{Erase}$^*$: Erase $n$ elements of an $n$ element container via an iterator.
\end{itemize}
Note that (\textsc{Access}$^*$) also appears in the Immer paper~\cite{puente17immer}.
That said, since Immer only supports {\em const} iterators, it cannot be evaluated on (\textsc{Append}$^*$), (\textsc{Update}$^*$), nor (\textsc{Erase}$^*$), since these micro-benchmarks require updates via an iterator.
Note also that the (\textsc{Concat}) benchmark, from~\autoref{sec:containers}, has been replaced by (\textsc{Erase}$^*$) since (1) the iterator version of (\textsc{Concat}) is similar to (\textsc{Append}$^*$), and (2) to test erasure via an iterator.

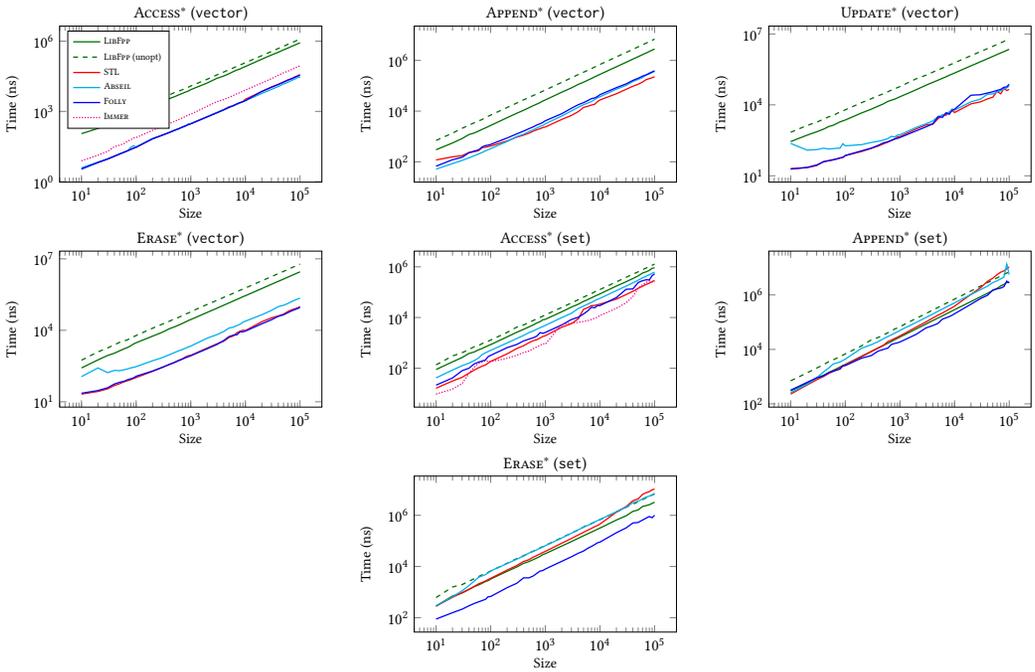
\begin{figure}
\centering
\begin{tikzpicture}[scale=0.6]
\begin{axis}[
    width=7.4cm,
    height=5cm,
    xlabel={Size},
    ylabel={Time (ns)},
    title={{\sc Access}$^*$ ({\tt vector})},
    title style={yshift=-1.5ex},
    grid=none,
    grid style={white!30!black},
    legend style={
        at={(0.03,0.97)},
        anchor=north west,
        draw=none,
        fill=none,
        text=black,
        font=\tiny,
        cells={anchor=west},
        draw=black,
        fill=white,
    },
    ticklabel style={text=black},
    label style={font=\small, yshift=0.5ex, text=black},
    title style={text=black},
    axis background/.style={fill=white},
    every axis plot/.append style={thick},
    xmode=log,
    ymode=log,
]

\addplot[
    color=darkgreen,
] table [x=Size, y=F::vector, col sep=comma] {data/access-it-vector.csv};

\addlegendentry{\tool}

\addplot[
    color=darkgreen, dashed,
] table [x=Size, y=F::vector, col sep=comma] {data/access-it-vector-slow.csv};

\addlegendentry{\tool (unopt)}

\addplot[
    color=red,
] table [x=Size, y=std::vector, col sep=comma] {data/access-it-vector.csv};
\addlegendentry{STL}

\addplot[
    color=cyan,
] table [x=Size, y=absl::InlinedVector, col sep=comma] {data/access-it-vector.csv};
\addlegendentry{\sc Abseil}

\addplot[
    color=blue,
] table [x=Size, y=folly::fbvector, col sep=comma] {data/access-it-vector.csv};
\addlegendentry{\sc Folly}

\addplot[
    color=magenta, densely dotted,
] table [x=Size, y=immer::flex_vector, col sep=comma] {data/access-it-vector.csv};
\addlegendentry{\sc Immer}

\end{axis}
\end{tikzpicture}
\quad
\begin{tikzpicture}[scale=0.6]
\begin{axis}[
    width=7.4cm,
    height=5cm,
    xlabel={Size},
    ylabel={Time (ns)},
    title={{\sc Append}$^*$ ({\tt vector})},
    title style={yshift=-1.5ex},
    grid=none,
    grid style={white!30!black},
    legend style={
        at={(0.03,0.97)},
        anchor=north west,
        draw=none,
        fill=none,
        text=black,
        font=\footnotesize,
        cells={anchor=west},
        draw=black,
        fill=white,
    },
    ticklabel style={text=black},
    label style={font=\small, yshift=0.5ex, text=black},
    title style={text=black},
    axis background/.style={fill=white},
    every axis plot/.append style={thick},
    xmode=log,
    ymode=log,
]

\addplot[
    color=darkgreen,
] table [x=Size, y=F::vector, col sep=comma] {data/append-it-vector.csv};

\addplot[
    color=darkgreen, dashed,
] table [x=Size, y=F::vector, col sep=comma] {data/append-it-vector-slow.csv};

\addplot[
    color=red,
] table [x=Size, y=std::vector, col sep=comma] {data/append-it-vector.csv};

\addplot[
    color=cyan,
] table [x=Size, y=absl::InlinedVector, col sep=comma] {data/append-it-vector.csv};

\addplot[
    color=blue,
] table [x=Size, y=folly::fbvector, col sep=comma] {data/append-it-vector.csv};

\end{axis}
\end{tikzpicture}
\quad
\begin{tikzpicture}[scale=0.6]
\begin{axis}[
    width=7.4cm,
    height=5cm,
    xlabel={Size},
    ylabel={Time (ns)},
    title={{\sc Update}$^*$ ({\tt vector})},
    title style={yshift=-1.5ex},
    grid=none,
    grid style={white!30!black},
    legend style={
        at={(0.03,0.97)},
        anchor=north west,
        draw=none,
        fill=none,
        text=black,
        font=\footnotesize,
        cells={anchor=west},
        draw=black,
        fill=white,
    },
    ticklabel style={text=black},
    label style={font=\small, yshift=0.5ex, text=black},
    title style={text=black},
    axis background/.style={fill=white},
    every axis plot/.append style={thick},
    xmode=log,
    ymode=log,
]

\addplot[
    color=darkgreen,
] table [x=Size, y=F::vector, col sep=comma] {data/update-it-vector.csv};

\addplot[
    color=darkgreen, dashed,
] table [x=Size, y=F::vector, col sep=comma] {data/update-it-vector-slow.csv};

\addplot[
    color=red,
] table [x=Size, y=std::vector, col sep=comma] {data/update-it-vector.csv};

\addplot[
    color=cyan,
] table [x=Size, y=absl::InlinedVector, col sep=comma] {data/update-it-vector.csv};

\addplot[
    color=blue,
] table [x=Size, y=folly::fbvector, col sep=comma] {data/update-it-vector.csv};

\end{axis}
\end{tikzpicture}

\begin{tikzpicture}[scale=0.6]
\begin{axis}[
    width=7.4cm,
    height=5cm,
    xlabel={Size},
    ylabel={Time (ns)},
    title={{\sc Erase}$^*$ ({\tt vector})},
    title style={yshift=-1.5ex},
    grid=none,
    grid style={white!30!black},
    legend style={
        at={(0.03,0.97)},
        anchor=north west,
        draw=none,
        fill=none,
        text=black,
        font=\footnotesize,
        cells={anchor=west},
        draw=black,
        fill=white,
    },
    ticklabel style={text=black},
    label style={font=\small, yshift=0.5ex, text=black},
    title style={text=black},
    axis background/.style={fill=white},
    every axis plot/.append style={thick},
    xmode=log,
    ymode=log,
]

\addplot[
    color=darkgreen,
] table [x=Size, y=F::vector, col sep=comma] {data/erase-it-vector.csv};

\addplot[
    color=darkgreen, dashed,
] table [x=Size, y=F::vector, col sep=comma] {data/erase-it-vector-slow.csv};

\addplot[
    color=red,
] table [x=Size, y=std::vector, col sep=comma] {data/erase-it-vector.csv};

\addplot[
    color=cyan,
] table [x=Size, y=absl::InlinedVector, col sep=comma] {data/erase-it-vector.csv};

\addplot[
    color=blue,
] table [x=Size, y=folly::fbvector, col sep=comma] {data/erase-it-vector.csv};

\end{axis}
\end{tikzpicture}
\quad
\begin{tikzpicture}[scale=0.6]
\begin{axis}[
    width=7.4cm,
    height=5cm,
    xlabel={Size},
    ylabel={Time (ns)},
    title={{\sc Access}$^*$ ({\tt set})},
    title style={yshift=-1.5ex},
    grid=none,
    grid style={white!30!black},
    legend style={
        at={(0.03,0.97)},
        anchor=north west,
        draw=none,
        fill=none,
        text=black,
        font=\footnotesize,
        cells={anchor=west},
        draw=black,
        fill=white,
    },
    ticklabel style={text=black},
    label style={font=\small, yshift=0.5ex, text=black},
    title style={text=black},
    axis background/.style={fill=white},
    every axis plot/.append style={thick},
    xmode=log,
    ymode=log,
]

\addplot[
    color=darkgreen,
] table [x=Size, y=F::set, col sep=comma] {data/access-it-set.csv};

\addplot[
    color=darkgreen, dashed,
] table [x=Size, y=F::set, col sep=comma] {data/access-it-set-slow.csv};

\addplot[
    color=red,
] table [x=Size, y=std::set, col sep=comma] {data/access-it-set.csv};

\addplot[
    color=cyan,
] table [x=Size, y=absl::btree_set, col sep=comma] {data/access-it-set.csv};

\addplot[
    color=blue,
] table [x=Size, y=folly::F14FastSet, col sep=comma] {data/access-it-set.csv};

\addplot[
    color=magenta, densely dotted,
] table [x=Size, y=immer::set, col sep=comma] {data/access-it-set.csv};

\end{axis}
\end{tikzpicture}
\quad
\begin{tikzpicture}[scale=0.6]
\begin{axis}[
    width=7.4cm,
    height=5cm,
    xlabel={Size},
    ylabel={Time (ns)},
    title={{\sc Append}$^*$ ({\tt set})},
    title style={yshift=-1.5ex},
    grid=none,
    grid style={white!30!black},
    legend style={
        at={(0.03,0.97)},
        anchor=north west,
        draw=none,
        fill=none,
        text=black,
        font=\footnotesize,
        cells={anchor=west},
        draw=black,
        fill=white,
    },
    ticklabel style={text=black},
    label style={font=\small, yshift=0.5ex, text=black},
    title style={text=black},
    axis background/.style={fill=white},
    every axis plot/.append style={thick},
    xmode=log,
    ymode=log,
]

\addplot[
    color=darkgreen,
] table [x=Size, y=F::set, col sep=comma] {data/append-it-set.csv};

\addplot[
    color=darkgreen, dashed,
] table [x=Size, y=F::set, col sep=comma] {data/append-it-set-slow.csv};

\addplot[
    color=red,
] table [x=Size, y=std::set, col sep=comma] {data/append-it-set.csv};

\addplot[
    color=cyan,
] table [x=Size, y=absl::btree_set, col sep=comma] {data/append-it-set.csv};

\addplot[
    color=blue,
] table [x=Size, y=folly::F14FastSet, col sep=comma] {data/append-it-set.csv};

\end{axis}
\end{tikzpicture}

\begin{tikzpicture}[scale=0.6]
\begin{axis}[
    width=7.4cm,
    height=5cm,
    xlabel={Size},
    ylabel={Time (ns)},
    title={{\sc Erase}$^*$ ({\tt set})},
    title style={yshift=-1.5ex},
    grid=none,
    grid style={white!30!black},
    legend style={
        at={(0.03,0.97)},
        anchor=north west,
        draw=none,
        fill=none,
        text=black,
        font=\footnotesize,
        cells={anchor=west},
        draw=black,
        fill=white,
    },
    ticklabel style={text=black},
    label style={font=\small, yshift=0.5ex, text=black},
    title style={text=black},
    axis background/.style={fill=white},
    every axis plot/.append style={thick},
    xmode=log,
    ymode=log,
]

\addplot[
    color=darkgreen,
] table [x=Size, y=F::set, col sep=comma] {data/erase-it-set.csv};

\addplot[
    color=darkgreen, dashed,
] table [x=Size, y=F::set, col sep=comma] {data/erase-it-set-slow.csv};

\addplot[
    color=red,
] table [x=Size, y=std::set, col sep=comma] {data/erase-it-set.csv};

\addplot[
    color=cyan,
] table [x=Size, y=absl::btree_set, col sep=comma] {data/erase-it-set.csv};

\addplot[
    color=blue,
] table [x=Size, y=folly::F14FastSet, col sep=comma] {data/erase-it-set.csv};

\end{axis}
\end{tikzpicture}

\caption{Figure showing iterator micro-benchmark results for {\em vectors} and {\em sets} in \tool and other libraries under different container sizes.  Scales are logarithmic.
Note that Immer's const-iterators can only be applied to \textsc{Access}$^*$.
Compared to the STL baseline, \tool is $23.6{\times}$/$11.8{\times}$/$45.2{\times}$/$28.2{\times}$ slower for the \textsc{Access}$^*$/\textsc{Append}$^*$/\textsc{Update}$^*$/\textsc{Erase}$^*$ (\texttt{vector}) benchmarks,
$3.1{\times}$ slower for the \textsc{Access}$^*$ (\texttt{set}) benchmark, and $2.9{\times}$/$2.9{\times}$ faster for the \textsc{Append}$^*$/\-\textsc{Erase}$^*$ (\texttt{set}) benchmarks.
\label{fig:vector-it-bench}}
\end{figure}

\myparagraph{Results}
The iterator-focused micro-benchmark results are summarized in \autoref{fig:vector-it-bench}.
In sequential containers, \tool has the same algorithmic complexity trend as other popular libraries, but requires a larger constant factor in access-frequent benchmarks due to the inherent internal overhead of persistent iterators.
This is not surprising, since iterator operations like \verb_++it_ compile down into a few instructions for array-based container implementations, but require a {\em zipper move} operation for persistent iterators.
The overheads are much narrower for node-based containers, reflecting the inherent cost of node traversal.
The tradeoff is a much safer usage of iterators with value semantics, and many operations (such as insertion and deletion) carry no risk of iterator invalidation or other aliasing hazards.
\tool also showed competitive performance advantages when employing modification-heavy workloads such as (\textsc{Erase}$^*$).
Comparison with a slower version of \tool also shows that the allocator choice manifested as a constant overhead independent of container size, and thus does not affect the complexity trend of the library.
Finally, Immer, with only support for \emph{const} iterators, is unable to execute all of the micro-benchmarks.

\result{\tool achieves practical performance comparable to existing persistent libraries, while offering a more general and expressive persistent iterator design.\vspace{0.5em}}

\subsection{Memory Overheads} \label{sec:rq3}

\begin{figure}[h]
\centering
\begin{tikzpicture}[scale=0.6]
\begin{axis}[
    width=7.4cm,
    height=5cm,
    xlabel={Size},
    ylabel={Allocated (bytes)},
    title={\tt vector},
    title style={yshift=-1.5ex},
    grid=none,
    grid style={white!30!black},
    legend style={
        at={(0.03,0.97)},
        anchor=north west,
        draw=none,
        fill=none,
        text=black,
        font=\footnotesize,
        cells={anchor=west},
        draw=black,
        fill=white,
    },
    ticklabel style={text=black},
    label style={font=\small, yshift=0.5ex, text=black},
    title style={text=black},
    axis background/.style={fill=white},
    every axis plot/.append style={thick},
    xmode=log,
    ymode=log,
]

\addplot[
    color=darkgreen,
] table [x=Size, y=F::vector, col sep=comma] {data/vector-size.csv};

\addlegendentry{\tool}

\addplot[
    color=red,
] table [x=Size, y=std::vector, col sep=comma] {data/vector-size.csv};
\addlegendentry{STL}

\addplot[
    color=cyan,
] table [x=Size, y=absl::InlinedVector, col sep=comma] {data/vector-size.csv};
\addlegendentry{\sc Abseil}

\addplot[
    color=blue,
] table [x=Size, y=folly::fbvector, col sep=comma] {data/vector-size.csv};
\addlegendentry{\sc Folly}

\addplot[
    color=magenta, densely dotted,
] table [x=Size, y=immer::flex_vector, col sep=comma] {data/vector-size.csv};
\addlegendentry{\sc Immer}

\end{axis}
\end{tikzpicture}
\qquad
\begin{tikzpicture}[scale=0.6]
\begin{axis}[
    width=7.4cm,
    height=5cm,
    xlabel={Size},
    ylabel={Allocated (bytes)},
    title={\tt set},
    title style={yshift=-1.5ex},
    grid=none,
    grid style={white!30!black},
    legend style={
        at={(0.03,0.97)},
        anchor=north west,
        draw=none,
        fill=none,
        text=black,
        font=\footnotesize,
        cells={anchor=west},
        draw=black,
        fill=white,
    },
    ticklabel style={text=black},
    label style={font=\small, yshift=0.5ex, text=black},
    title style={text=black},
    axis background/.style={fill=white},
    every axis plot/.append style={thick},
    xmode=log,
    ymode=log,
]

\addplot[
    color=darkgreen,
] table [x=Size, y=F::set, col sep=comma] {data/set-size.csv};

\addplot[
    color=red,
] table [x=Size, y=std::set, col sep=comma] {data/set-size.csv};

\addplot[
    color=cyan,
] table [x=Size, y=absl::btree_set, col sep=comma] {data/set-size.csv};

\addplot[
    color=blue,
] table [x=Size, y=folly::F14FastSet, col sep=comma] {data/set-size.csv};

\addplot[
    color=magenta, densely dotted,
] table [x=Size, y=immer::set, col sep=comma] {data/set-size.csv};

\end{axis}
\end{tikzpicture}
\caption{Figure showing heap consumption for {\em vectors} and {\em sets} in \tool and other libraries under different container size.
Scales are logarithmic. For Immer, transient-mode is not used.\label{fig:heap-alloc}}
\end{figure}
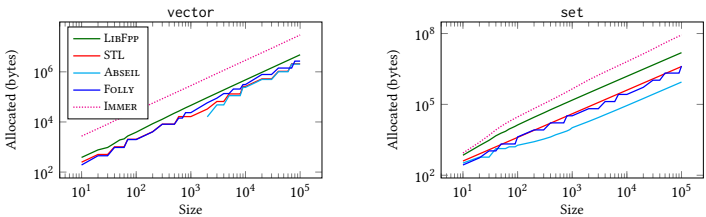

\noindent
In addition to the asymptotic complexity and run time of containers, memory consumption is also an important metric.
In light of this, each of \tool and other baseline containers are parameterized with unsigned 64-bit integers, and their heap consumption with regard to different numbers of elements stored is summarized in \autoref{fig:heap-alloc}.

Due to the more complex data structure hidden within \tool and Immer's containers, their respective heap memory consumption is higher than that of comparatively simple containers, such as \texttt{std::vector}.
However, as the scale of the dataset grows, \tool's provided vectors and sets can still attain comparable memory consumption compared to baseline offerings, proving its scalability across a wide range of data sizes.
One notable thing from the figure is the behavior of \verb_absl::InlinedVector_, which by default stores the first $10^3$ elements on the stack for better performance, and thus will have 0 heap consumption when using smaller sizes.

\result{\tool achieves reasonable memory consumption comparable to existing baseline libraries, while further reducing its memory overhead in real workflows due to the usage of copy-on-write-like techniques that avoid copying until necessary.\vspace{0.5em}}

\section{Related Work}

\myparagraph{Functional and Declarative Languages}
Functional and declarative languages (Haskell, ML, and their derivatives) support persistent data structures as a natural consequence of immutability.
However, these languages generally do not provide iterator-like abstractions similar to \verb_C++_.
Rather, traversal in these languages is typically achieved via higher-order {\em combinators} (e.g., \verb+map+, \verb+foldl+, \verb+zipWith+) or recursion.
However, as noted in \autoref{sec:example}, these have limitations in natural expressiveness, abstraction, or performance.

\myparagraph{Zippers}
{\em Zipper}s have been explored in functional programming literature and practice---e.g., from Huet’s original paper~\cite{HUET1997}, to later extensions for trees, syntax manipulation, and XML editing.
However, zippers are rarely used in mainstream standard libraries for functional languages, and mostly appear as third-party packages (e.g., \verb+Data.Tree.Zipper+), if at all.
Similarly, ML-family languages do not include zipper-based traversal in their core persistent collection libraries.
The problem with zippers in functional/declarative languages is two-fold.
Firstly, each movement constructs a new path context, generating transient allocations and GC pressure.
Secondly, zippers introduce a {\em stateful} cursor abstraction that tends to feel unnatural in declarative languages.
\tool revisits the zipper concept in an imperative setting where iterators are familiar and idiomatic, and performance is addressed through custom memory management and optimizations such as destructive update.
As far as we are aware, \tool is the first implementation of zippers for finger trees and the first to use zippers to implement an STL-like iterator-like API.

\myparagraph{Immer}
\textit{Immer}~\cite{puente17immer} is a modern \verb_C++_ library providing persistent containers, and is perhaps the closest existing system to our work.
Immer implements sequence containers using RRB-trees~\cite{stucki15rrb} and associative containers using CHAMPs~\cite{steindorfer15champ}.
Immer uses a large node fan-out for performance, allowing for fast random $O(\log_{32} N)$ access while also preserving structural sharing.
In contrast, our design prioritizes a uniform abstraction over recovering the constant factors of mutable data structures.
Immer’s public iterators are {\em read-only}: i.e., \verb+iterator+ and \verb+const_iterator+ alias to the same type.
In contrast, \tool provides full {\em persistent} iterators with value semantics, allowing for updates (to a local copy) via the iterator.
Finally, Immer exposes {\em persistence} and {\em transience} through explicit modes, allowing developers to manually switch between the two for performance tuning.
In contrast, \tool is {\em transparent}: persistence is automatic and orthogonal to syntax, supporting familiar STL-style while also benefiting from efficient structural sharing. 

\myparagraph{Abseil, Folly, and Boost}
Several widely used \verb_C++_ libraries extend or complement the STL, including Abseil~\cite{Abseil} (Google), Folly~\cite{Folly} (Meta), and Boost~\cite{boost}.
These frameworks share a common goal of improving the performance, safety, and ergonomics of standard containers, but generally do not attempt to provide persistent or immutable data structures.
\tool is also intended to complement the STL, by providing persistent versions of containers/iterators while also supporting STL-like patterns and idioms. 

\section{Conclusion}

Persistent data structures are often perceived as elegant but impractical for systems programming. 
This work demonstrates that, with careful engineering, they can be made practical and ergonomic within the expected trade-offs of persistence. 
We introduced \emph{persistent iterators}, a design that reconciles the iterator-based programming model of imperative languages with the immutability guarantees of persistent data structures. 
By combining finger trees with zipper-based traversal, \tool provides a uniform, STL-like 
interface supporting structural sharing, persistence, and STL-style iterator operations under value semantics; thereby eliminating invalidation and aliasing hazards.

Our analysis and evaluation show that \tool achieves asymptotic complexities comparable to standard STL containers, albeit with higher constant factors consistent with persistence.
We show these additional overheads can be reduced through a combination of custom memory management, destructive update optimizations, and cache-aware node layouts.
The resulting library makes persistence and pure value semantics more accessible under traditional (imperative, mutation-based) programming styles, and can serve as a practical alternative for modern software where persistence and immutability are desired.
By bringing persistence and value semantics into the familiar idioms of iterator-based programming, \tool enables developers to write safer, side-effect-free, and conceptually simpler code.

\subsection*{Acknowledgements}

This research is supported by the National Research Foundation, Singapore, under its National Cybersecurity R\&D Programme (Award No. CRPO-GC5-NUS-004).

\newpage

\bibliographystyle{ACM-Reference-Format}
\bibliography{main}

\end{document}